\documentclass[intlimits,twoside,a4paper]{article}

\usepackage[cp1251]{inputenc}

\usepackage[eqsecnum]{cmpj3}

\issue{2025}{28}{4}{43703}
\doinumber{10.5488/CMP.28.43703}

\title[A geometric analysis near the phase transitions in Sn$_2$P$_2$S$_6$]%
{Shape descriptors of equilibrium states in a quantum lattice model with local multi-well potentials: A geometric analysis near the phase transitions in Sn$_2$P$_2$S$_6$ ferroelectric crystals}
\author[S.  \"{O}z\"{u}m, T. Akkurt, R. Erdem, N. G\"{u}\c{c}l\"{u}]
{S. \"{O}z\"{u}m\orcid{0000-0003-2123-5856}\refaddr{label1},
T. Akkurt \orcid{0000-0002-9107-443X}\refaddr{label2},
R. Erdem \orcid{0000-0001-7370-2263}\refaddr{label3}\thanks{Corresponding author: \email{rerdem@akdeniz.edu.tr}},
N. G\"{u}\c{c}l\"{u}\orcid{0000-0002-0069-0238}\refaddr{label4}}

\addresses{
\addr{label1} Alaca Avni \c{C}elik Vocational School, Hitit University, 19600  \c{C}orum, T\"{u}rkiye
\addr{label2} Institute of Science, Akdeniz University, 07058 Antalya, T\"{u}rkiye	
\addr{label3} Department of Physics, Akdeniz University, 07058 Antalya, T\"{u}rkiye
\addr{label4} Department of Physics Education, Necmettin Erbakan University,  42090 Konya,  T\"{u}rkiye}


\date{Received April 30, 2025, in final form September 02, 2025}

\begin{document}
	
\maketitle
	
	\begin{abstract}
We analyze the equilibrium states of quantum lattice model with local multi-well potentials for  Sn$_2$P$_2$S$_6$ ferroelectric crystals using the mean and Gaussian curvatures ($H$, $K$),  curvedness ($C$) and shape index ($S$).  From the energy gap, pressure and temperature variations of $H$, $K$, $C$ and $S$, we have reported the geometric construction of the free energy surfaces for the ferroelectric and paraelectric phases. Their behaviors are explicitly observed near the ferroelectric-paraelectric phase transitions. It is found that $H$, $C$ and $S$ display a cusp singularity at the criticality while $K$ converges to zero on both sides of the critical and tricritical points.
\printkeywords
\end{abstract}

\section{Introduction}
	
Shape descriptors, such as  mean and Gaussian curvatures, curvedness, shape index, surface area and volume are employed to describe the differential geometry \cite{1}. Especially, the nontrivial notions of curved geometry characterized by mean curvature ($H$), Gaussian curvature ($K$), curvedness ($C$) and shape index~($S$) are becoming prevalent and exist concomitantly in physical and biophysical systems with regard to many microscopic and macroscale objects \cite{2,3,4,5}. Materials science and engineering are definitely playing a central role in this endeavor through the signs of these quantities, where a face is classified into eight fundamental surface types represented as pit, valley, saddle valley, flat, minimal surface, saddle ridge, ridge and peak  \cite{6,7,8,9,10,11}. Moreover, new and  most important trend in mathematics and physics is the study of singularities in these geometric features, which plays an essential role in many physical events, such as changing topology or the emergence of a new structure. Therefore it is easy to determine the stable, metastable and unstable states in the free energy profile if one knows the signs of these descriptors. For example, the minimal surfaces are obtained when $H=0$ and $K<0$ \cite{11}.

The objective of the present paper is to provide an expository investigation of the above tools ($H$, $K$, $C$, $S$) for analysing a free energy surface and  geometric modelling of equilibrium states  in the quantum lattice model (QLM). In the last decade, many experimental and theoretical works have been devoted to the study of the quantum lattice model. In some of these studies, the  phase transitions between the ferroelectric (FE) and paraelectric (PE) phases in the  Sn$_2$P$_2$S$_6$ (SPS) crystals were presented by many authors \cite{12,13,14,15,16,17,18,19,20}. However, there has been no geometric analysis of the mean-field free energy surface using  ($H$, $K$, $C$, $S$). Different from the earlier investigation, where the phase transitions undergoing in the same system was first interpreted in the context of Ruppeiner geometry \cite{19},  our discussions fairly well describe some geometric features of equilibrium states by determining $H$, $K$, $C$ and $S$.

The rest of this paper is organized as follows. Section~\ref{sec-2} is devoted to theoretical model with a Hamiltonian equation and its mean-field equilibrium states. In order to analyze these states geometrically, in section~\ref{sec-3}, we mention the ways of calculating the mean and Gaussian curvatures, curvedness and  shape index numerically and show their energy gap $\varepsilon$, pressure $p$   and  temperature $\theta$ plots in section~\ref{sec-4}. This paper ends with concluding remarks in section~\ref{sec-Con}.

\section{The model and its equilibrium states}
\label{sec-2}
	
The Hamiltonian of the quantum lattice model is written in the form 
\begin{equation} \label{eq1}
	\hat{H}=\sum_i \hat{H}_1+\hat{H}_2+\hat{H}_3,
\end{equation}
\noindent which is composed of the single-site, interaction and deformation energy parts, respectively. These components have been described explicitly in references~\cite{18,19}.  For the Gibbs free energy of the system, it is easy to obtain, under the mean-field approximation (MFA), the following expression
\begin{equation} \label{eq2}
	\phi=\frac{G}{N}=\frac{1}{2} J \eta^2+\frac{1}{2} \nu c_0 u^2-\theta \ln \left[1+2 \exp \left(-\frac{\varepsilon+D u}{\theta}\right) \cosh \left(\frac{J \eta}{2 \theta}\right)\right]-\nu u \sigma_S,
\end{equation}
where $N$ is the number of lattice points, $J=\sum_{j}^{} J_{i j}$ is the effective field acting on dipoles, $\eta=\left\langle s_i\right\rangle$  is dipole ordering parameter ($s_i$ is variable related to the local dipole moment, $\left\langle...\right\rangle$ denotes the thermal expectation value), $\nu$  is volume related with one formula unit, $c_0$   is volume elastic constant,  $u$ is the deformation (or relative volume change), $\theta = kT$    is the reduced temperature ($k$ is Boltzmann constant, $T$ is absolute temperature),  $\varepsilon$ is the bare energy gap,  $D$ is the constant of an electron-deformational interaction and $\sigma_S = -p$   is the mechanical stress ($p$ is hydrostatic pressure). The equilibrium conditions $({\partial \phi}/{\partial \eta}=0$,  ${\partial \phi}/{\partial u}=0)$ result in the equations
\begin{equation}
	\eta=\frac{\re^{-b} c}{1+2 \re^{-b} a},
\end{equation}
\begin{equation} \label{eq3}
	\nu c_0+\frac{D}{\nu}\left(\frac{2 \re^{-b} a}{1+2 \re^{-b} a}\right)=\sigma,
\end{equation}
\noindent where $a=\cosh \left({J \eta}/{2 \theta}\right)$, $b={(\varepsilon+D u)}/{\theta}$, $c=\sinh \left({J \eta}/{2 \theta}\right)$. From the numerical solutions of the above equations, one can find the polarization  and deformation $(\eta, u)$   as functions of  $(\varepsilon, p, \theta)$. Since these  have been obtained for a choice of parameter values and discussed in detail in references~\cite{18,19}, we shall give a brief summary here as follows: In the $\eta$ vs. $\varepsilon$ and $u$ vs. $\varepsilon$ planes using $p=0$, for $\theta < 0.0179$~eV there is a first-order phase transition to a state with $\eta>0$, which corresponds to a FE phase, and for $\theta > 0.0179$~eV a transition of the second order exists to a state with  $\eta>0$ in the FE phase. The system has a tricritical point~(TCP) for $\theta=0.0179$~eV. Similarly, in the $\eta$ vs. $p$ and $u$ vs. $p$ plots based on  $\varepsilon=-0.011$~eV case, for temperatures less than $\theta_{\rm {TCP}}=0.0175$ eV a first-order phase transition occurs and that for  $\theta>\theta_{\rm {TCP}}$  a second-order phase transition exists.  This information is very important for studying the geometry of the free energy surfaces of the system.

\section{Definitions of shape descriptors and their meanings}
	\label{sec-3}

For the quantum lattice model, it is known that Gibbs free energy \eqref{eq2} which depends on the variables ($\eta$, $u$) describes a surface embedded in three-dimensional parameter space. Two principal curvatures ($\kappa_1, \kappa_2$) determine the local shape at a point on this surface. One characterizes the rate of maximum bending of the free energy surface and the tangent direction in which it occurs, while the other characterizes the rate and tangent direction of the minimum bending. Based on the standard facts from linear algebra,  ($\kappa_1, \kappa_2$) can be determined from the eigenvalues of the Hessian matrix of the free energy, with eigenvectors  ($e_1, e_2$). Since the Hessian is symmetric,  ($\kappa_1$, $\kappa_2$) are real and ($e_1$, $e_2$) are orthogonal. These are defined in terms of the mean ($H$) and Gaussian curvatures ($K$) as follows \cite{1,7, 21}
\begin{equation}
	\kappa_{1,2}=H \pm\left(H^2-K\right)^{1 / 2},
\end{equation}
\noindent where
\begin{equation} \label{eq4}
	H = \frac{(1+\phi_{2}^2)\phi_{11}-2\phi_1 \phi_2 \phi_{12}+(1+\phi_{1}^2)\phi_{22}}{2(1+\phi_1^2+ \phi_2^2)^{3/2}},
\end{equation}
\begin{equation} \label{eq5}
	K = \frac{\phi_{11}\phi_{22}-\phi_{12}^2}{(1+\phi_1^2+\phi_2^2)^2}.
\end{equation}
\noindent Here, the first- and second-order derivatives in equations \eqref{eq4} and \eqref{eq5}  are given as $\phi_1 = \partial \phi / \partial \eta$, $\phi_2 = \partial \phi / \partial u$, $\phi_{11} =  \partial \phi_{1} / \partial \eta$, $\phi_{12} =  \partial \phi_{1} / \partial u$, $\phi_{22} =  \partial \phi_{2} / \partial u$. On the other hand, the curvedness ($C$) and the shape index ($S$) were firstly proposed by Koenderink and van Doorn \cite{6} to describe a local surface type in terms of the principal curvatures ($\kappa_1$, $\kappa_2$). In their work, the curvedness is directly defined as
\begin{equation} \label{eq6}
		C = \sqrt{\frac{\kappa_1^2+\kappa_2^2}{2}},
\end{equation}
\noindent while the shape index, an important surface parameter complementary to the curvedness, is given by
\begin{equation} \label{eq7}
S = -\frac{2}{\piup} \arctan \frac{\kappa_1+\kappa_2}{\kappa_1-\kappa_2}.
\end{equation}
\noindent The $C$ value  always ranges from $0$ to $1$ while the  $S$ value can range between $-1$ and $+1$.

Each descriptor highlights different attributes of the underlying topology of the surface. The first feature is the general shape morphology and is defined by the various sign combinations of $\kappa_1$ and $\kappa_2$. The second feature is curvature magnitude, i.e., how bent the surface is irrespective of shape morphology.  The mean curvature  significantly differentiates the areas of high and low curvature, as well as convex and concave shapes. Gaussian curvature discriminates well between spherical and saddle-like areas. Curvedness reflects the absolute curvature magnitude in each point, irrespective of its specific shape. Finally, the shape index is capable of differentiating between pure shape characteristics, e.g., domes, ridges 
and saddles.
The curvature signs in these descriptors  mentioned above are summarized in table~\ref{table-1}.

\begin{table}[ht] 
	\centering
	\caption{Illustrations of $H$, $K$, $C$ and $S$ divided into eight categories. The different subintervals of $S$ for the interval $[-1,1]$ correspond to eight geometric surfaces \cite{22}.} 
	\label{table-1}
	\vspace{0.5em}
	\begin{tabular}{|l|l|l|l|l|} 
		\hline
		Surface Type & $H$ & $K$  & $C$ &  $S$   \\
		\hline
		Spherical cup/pit & $+$ &  $+$ &  $>0$ & $[-1,-0.625)$      \\
		\hline
		Valley/rut    & $+$ &  $0$ & $>0$ & $[-0.625,-0.375)$     \\
		\hline
		Saddle valley/rut  & $+$ &  $-$ & $>0$ & $[-0.375,-0.125)$    \\
		\hline
		Flat plane & $0$ & $0$  & $0$ &  $0$   \\
		\hline
		Saddle &   $0$ &  $-$  & $>0$ & $[-0.125,0.125)$   \\
		\hline
		Saddle ridge   & $-$ &  $-$ & $>0$ & $[0.125,0.375)$     \\
		\hline
		Ridge &   $-$ &  $0$ & $>0$ &$[0.375,0.625)$    \\
		\hline
		Spherical dome/peak  & $-$ &  $+$  & $>0$ & $[0.625,1)$    \\
		\hline

	\end{tabular}
	\label{3}
\end{table}

In the present work, we for the first time introduce the  above quantities to the free energy surfaces described by \eqref{eq2} in the quantum lattice model of ferroelectric crystals. They are evaluated in the equilibrium state and  expressed in terms of  $\eta$ and $u$. Then, from the numerical solutions of the self-consistent equations we easily find the ($H$, $K$, $C$, $S$) vs. $(\varepsilon, p, \theta)$ plots.

\section{Results and discussion}
\label{sec-4}

In the absence $(p=0)$ and presence $(p \neq 0)$ of hydrostatic pressure, a variety of critical and tricritical phenomena were investigated in the QLM model of ferroelectricity \cite{18}. Based on this investigation one of us (RE) studied the critical and tricritical properties of Ricci scalar (or thermodynamic curvature) \cite{19}. Similarly, here we have focused on energy gap, pressure and temperature dependencies of the mean and Gaussian curvatures as well as shape index and curvedness for the free energy surfaces at equilibrium.

\begin{figure*}[h]
	\centering
	\includegraphics[scale=0.25]{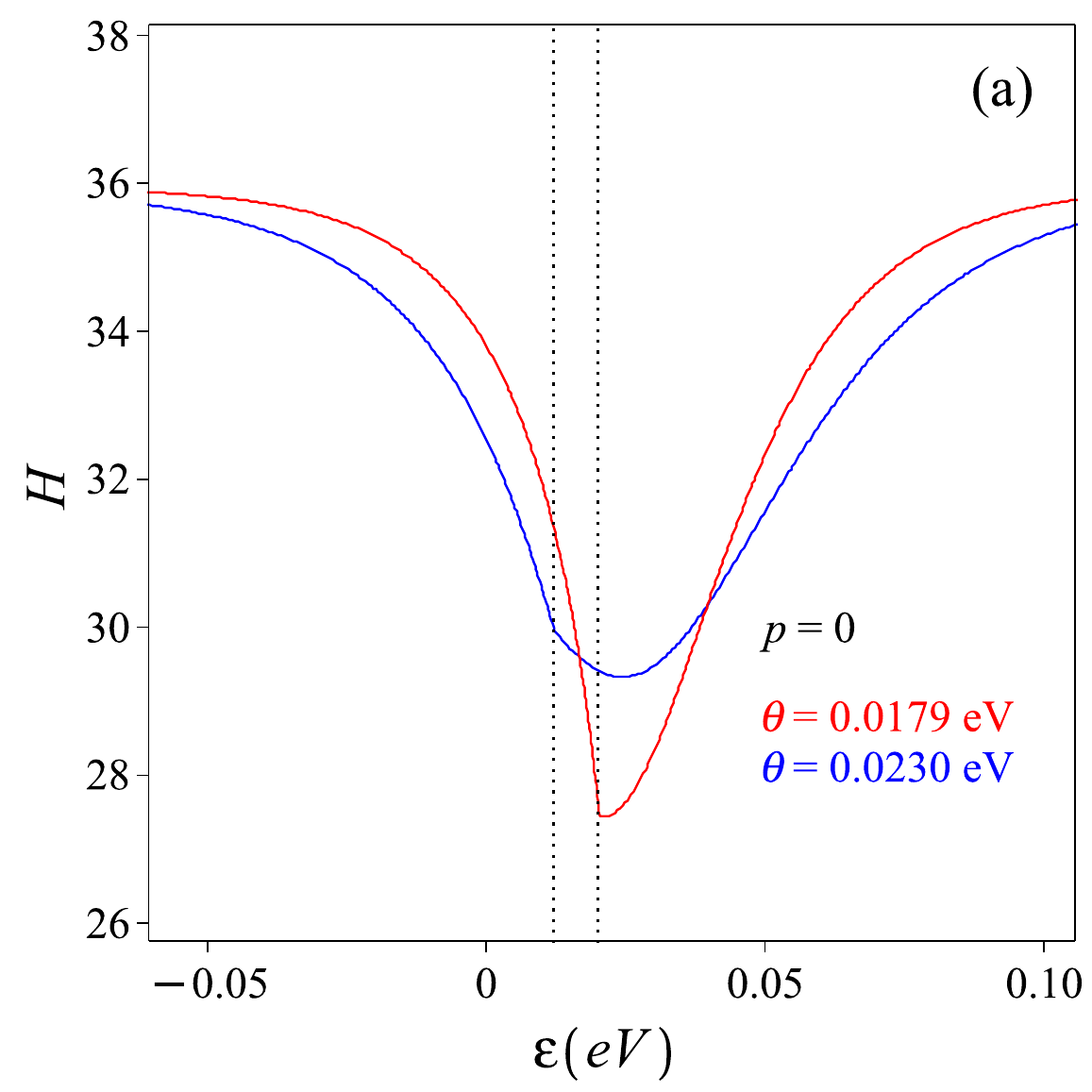} \label{1a}
	\includegraphics[scale=0.25]{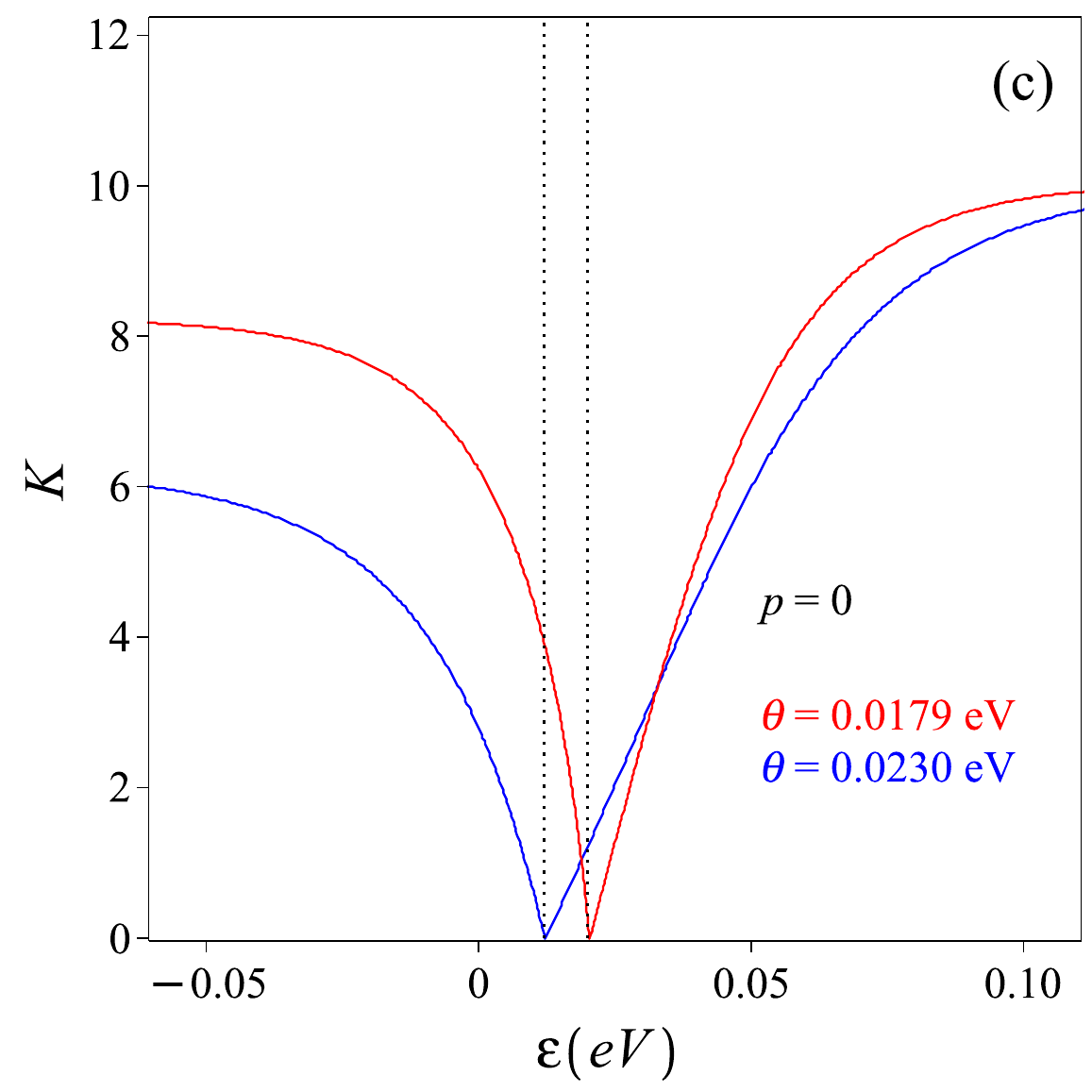}\\
	\includegraphics[scale=0.25]{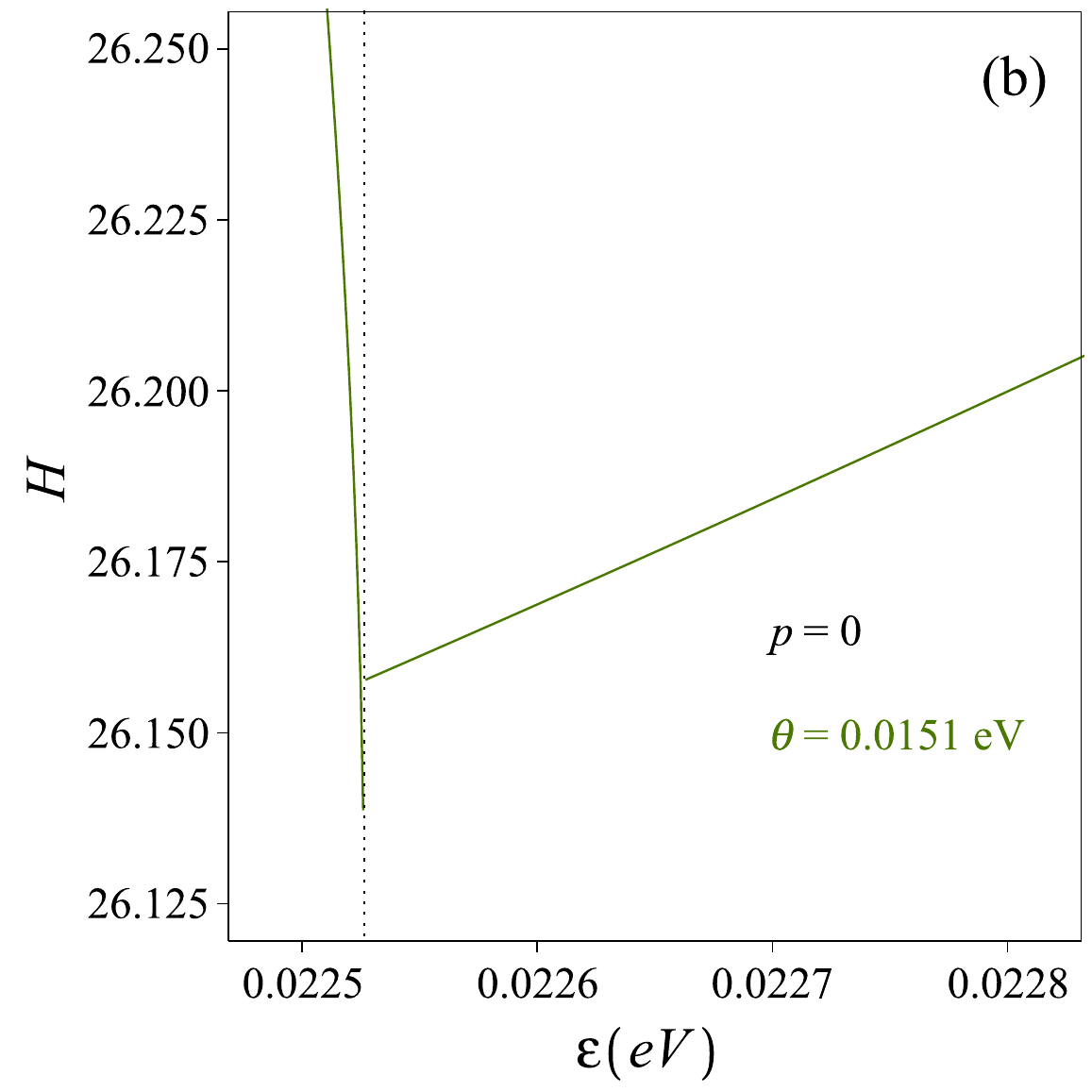}	\label{1c}
	\includegraphics[scale=0.25]{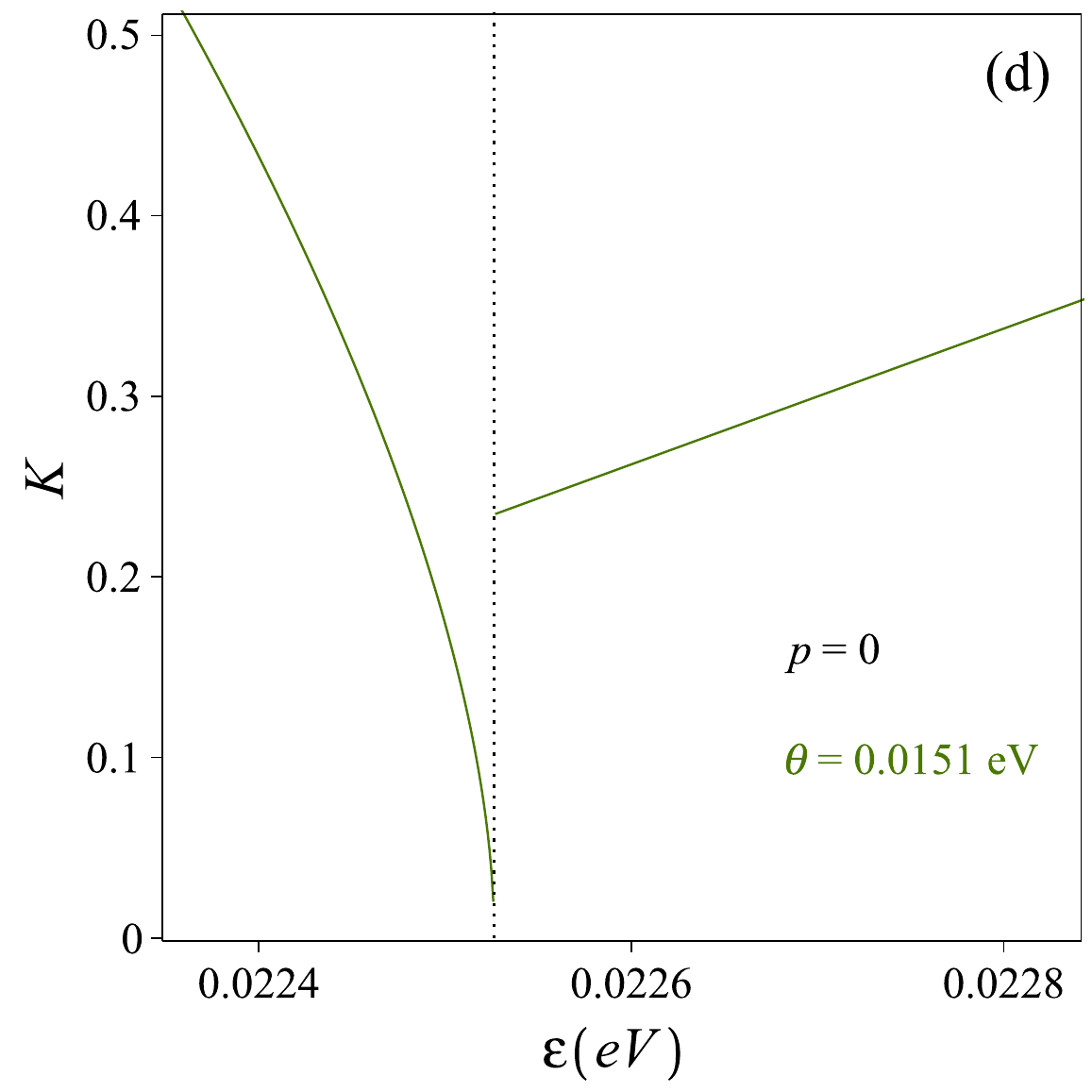}
	\caption{(Colour online) \textbf{(a)}, \textbf{(b)} $H$ and \textbf{(c)}, \textbf{(d)} $K$ vs. $\varepsilon$ for several temperatures  at $p=0$.}
	\label{fig1}
\end{figure*}

First, choosing $p=0$ and $\theta \leqslant 0.0230$ eV, we show typical mean and Gaussian curvatures vs. $\varepsilon$ near a  phase transitions from the FE to PE phase in figure~\ref{fig1}.
In the FE and PE phases $H>0$   and $K>0$  which corresponds to stable state and a “pit shape” in the free energy surface. As can be seen from the red and blue colored curves, there is no anomaly of $H$ similar to Ricci
scalar in reference~\cite{19} because $H$ is always finite and positive in the whole $\varepsilon$ range including $\varepsilon \geqslant \varepsilon_C$. As a critical behaviour
it only displays a cusp singularity at $\varepsilon_C$ ($\theta=0.0230$ eV)  and $\varepsilon_{\rm {TCP}}$  ($\theta=0.0179$ eV) [figure~\ref{fig1}(a)], while one observes a jump discontinuity of $H$  at the first-order
phase transition [figure~\ref{fig1}(b)]. Unlike $H$, the curvature $K$ (red and blue colored curves) converges to zero on both sides of the critical and tricritical $\varepsilon$ values, seen in figure~\ref{fig1}(c). From the findings in figures~ \ref{fig1}(a) and  \ref{fig1}(c) $(H>0$, $K \approx 0)$, one can note that the free energy surface becomes a ``valley shape'' during the continuous phase transition $(\varepsilon_C)$ and the tricritical point $(\varepsilon_{\rm {TCP}})$. One may also observe the same behaviour of $K$ versus $\varepsilon$ as $H$ vs. $\varepsilon$ at the first-order phase transition $(\varepsilon_t)$ in figure~\ref{fig1}(d). 

\begin{figure*}[h]
	\centering
	\includegraphics[scale=0.25]{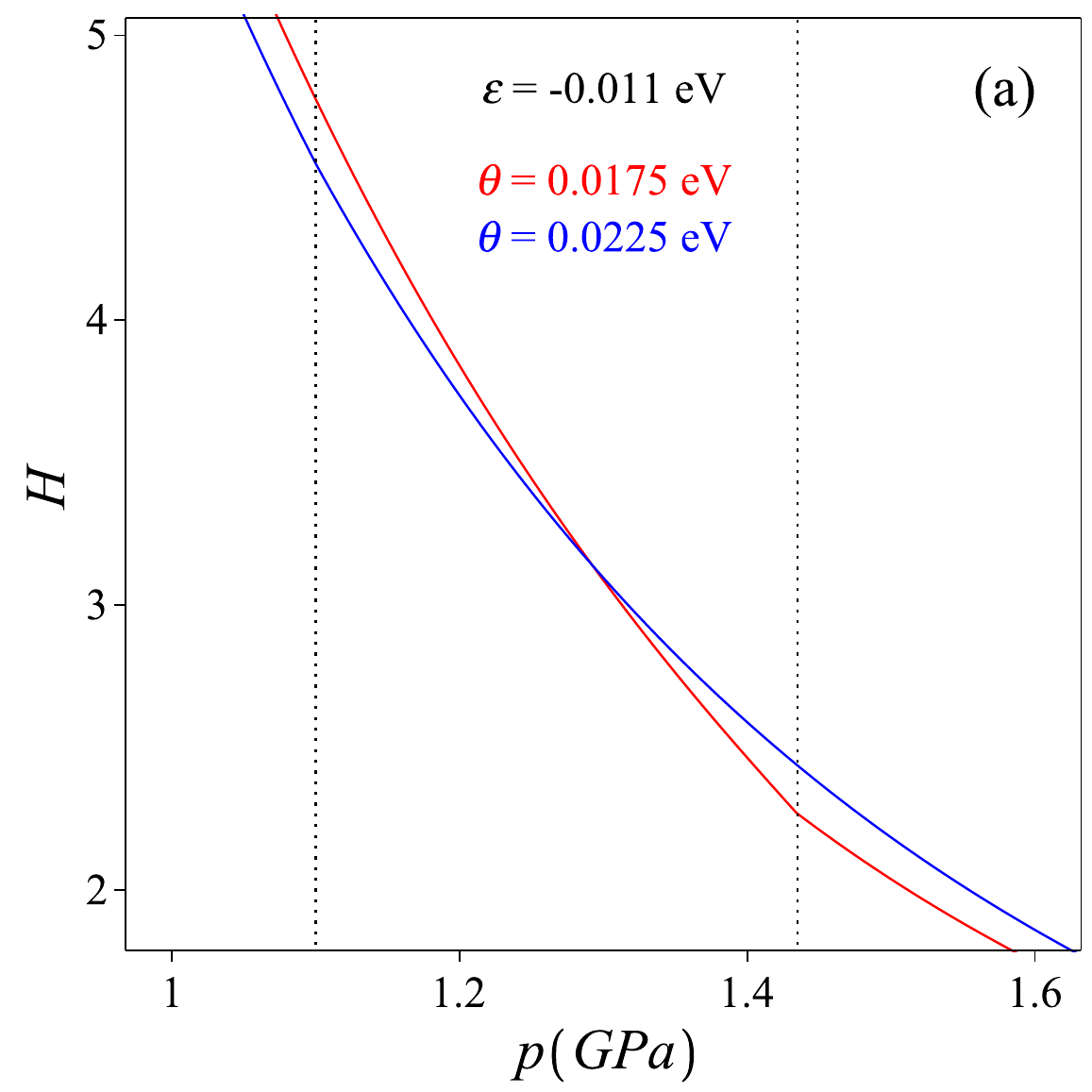}
	\includegraphics[scale=0.25]{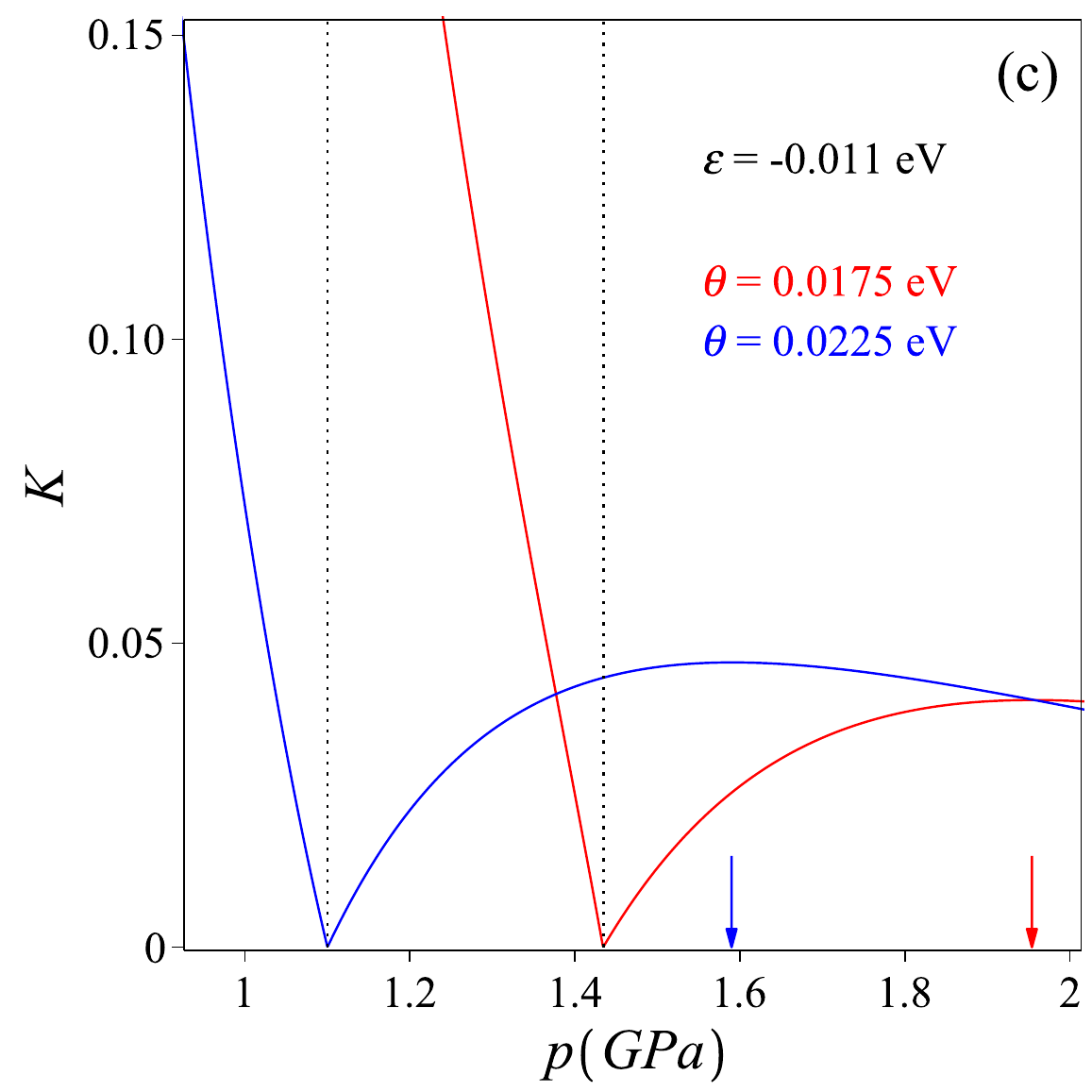} \\
	\includegraphics[scale=0.25]{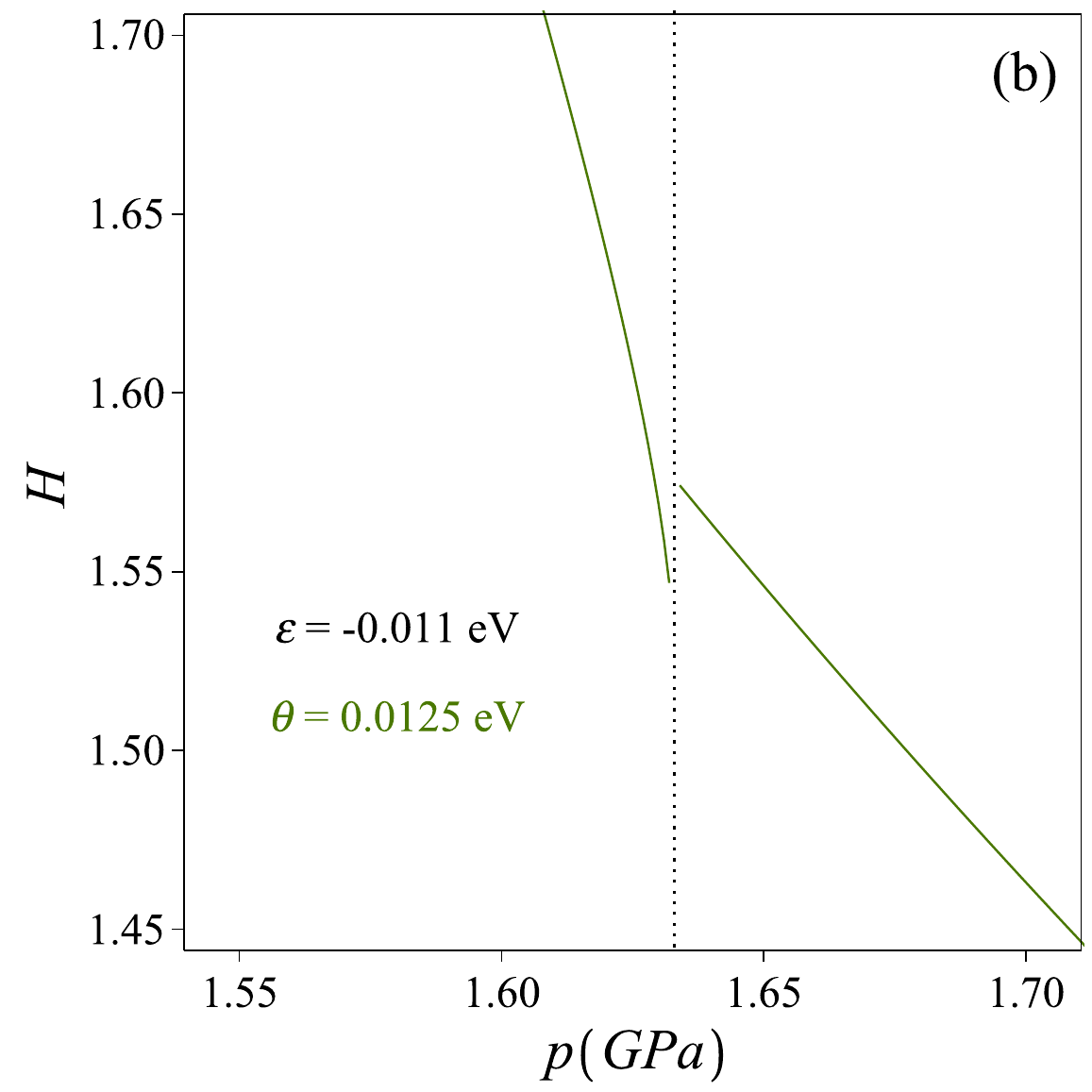}
	\includegraphics[scale=0.25]{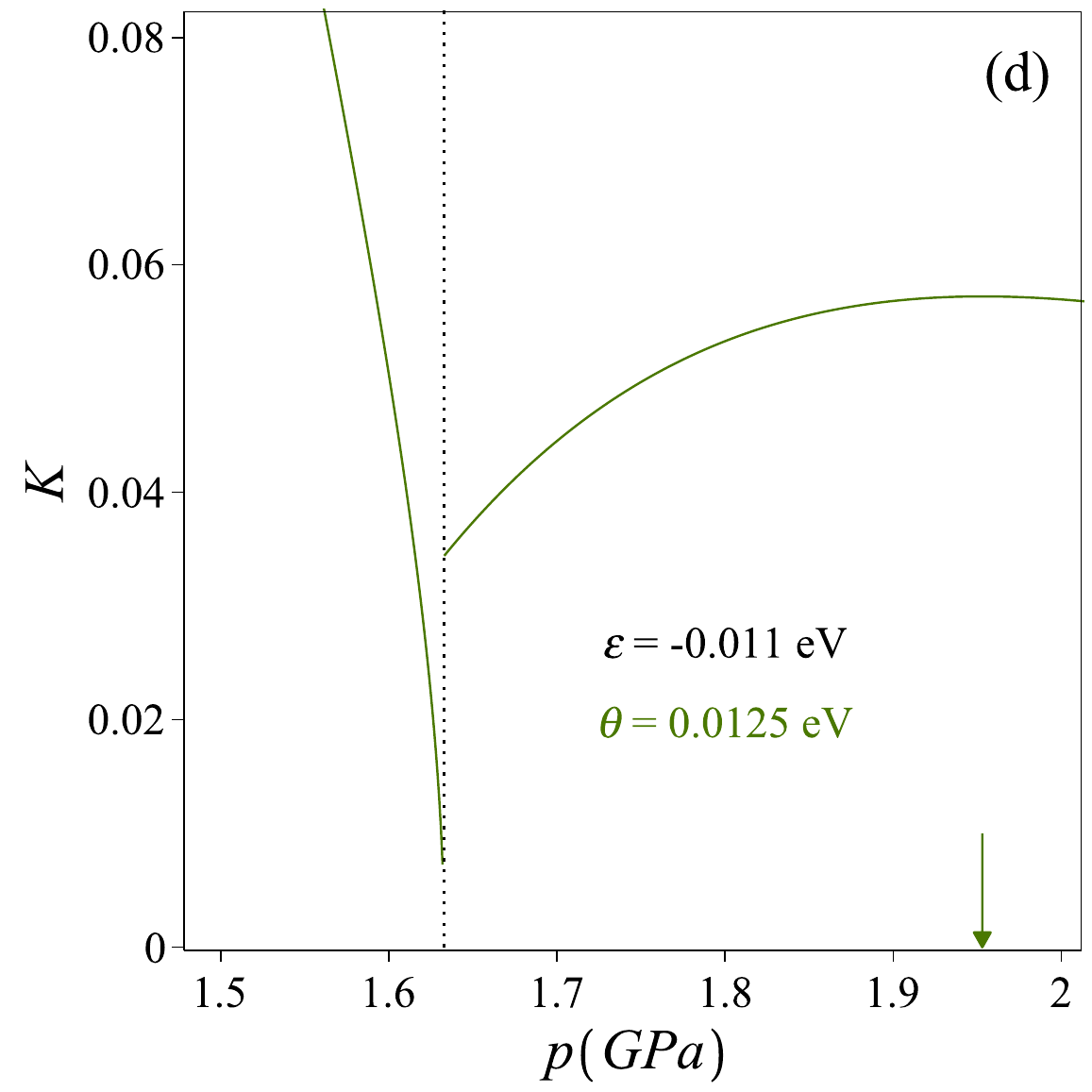}
	\caption{(Colour online) \textbf{(a)}, \textbf{(b)} $H$ and \textbf{(c)}, \textbf{(d)} $K$ vs. $p$ for several temperatures at $\varepsilon=-0.011$ eV.} \label{fig2}
\end{figure*}

\begin{figure*}[h!]
	\centering
	\includegraphics[scale=0.25]{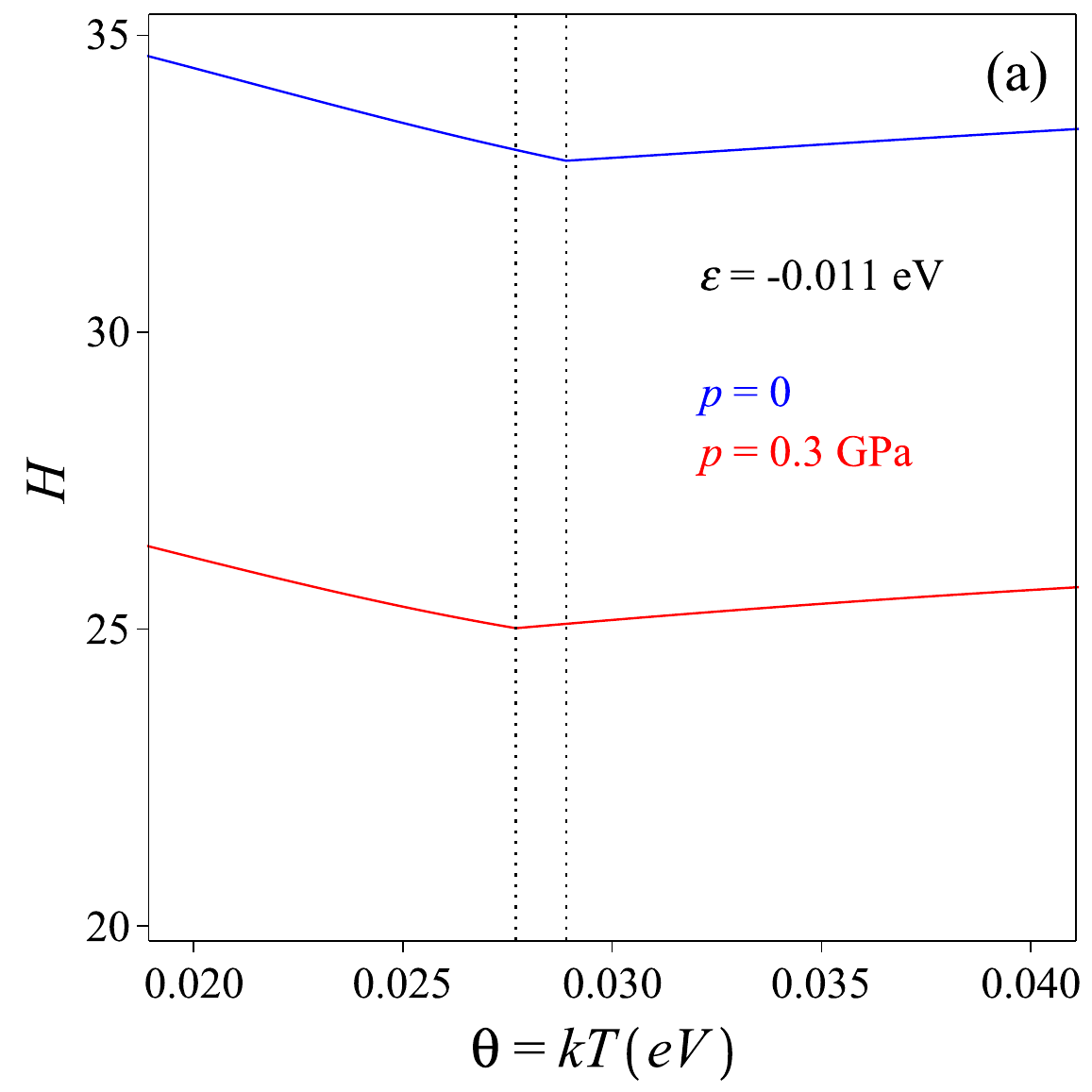}
	\includegraphics[scale=0.25]{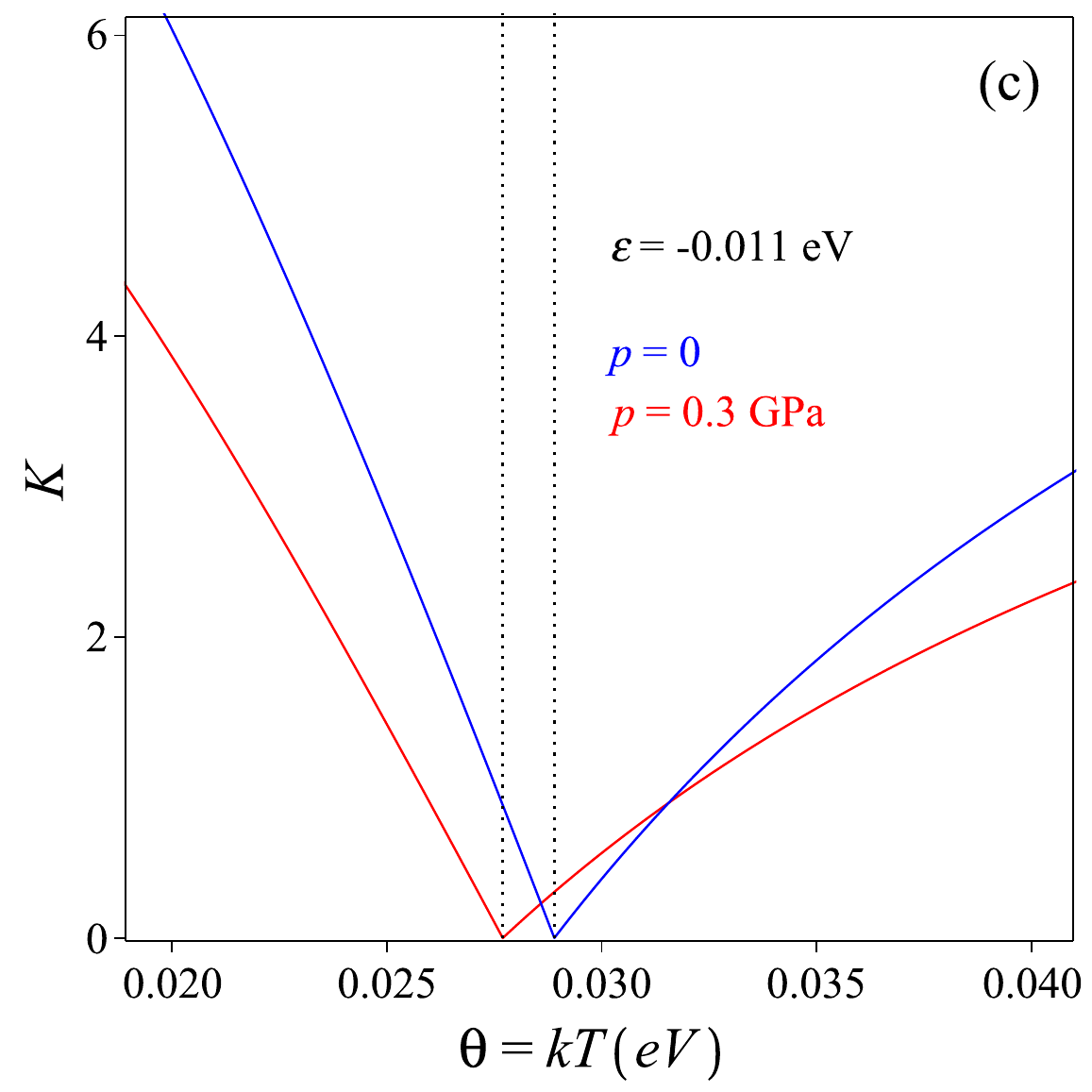} \\
	\includegraphics[scale=0.25]{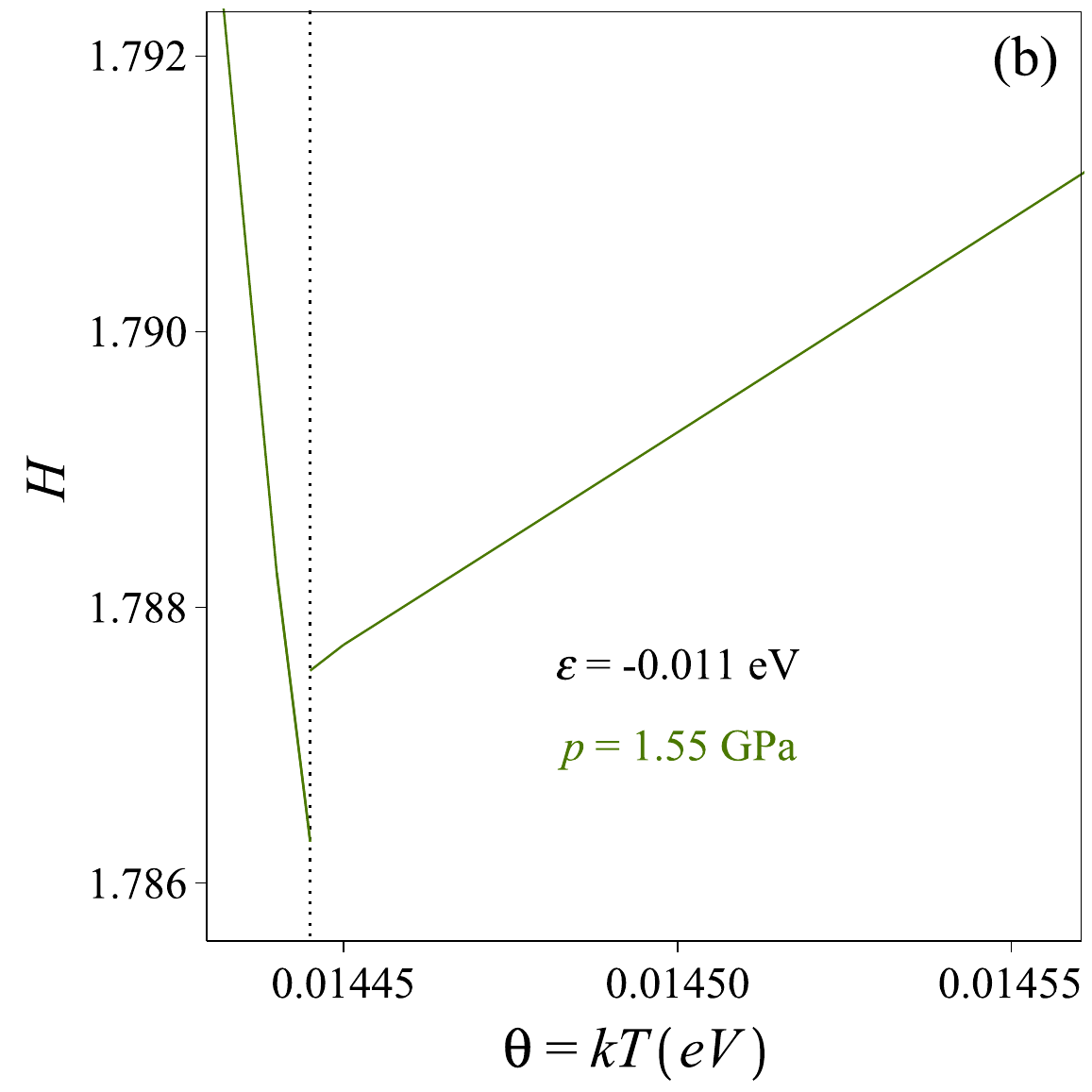}
	\includegraphics[scale=0.25]{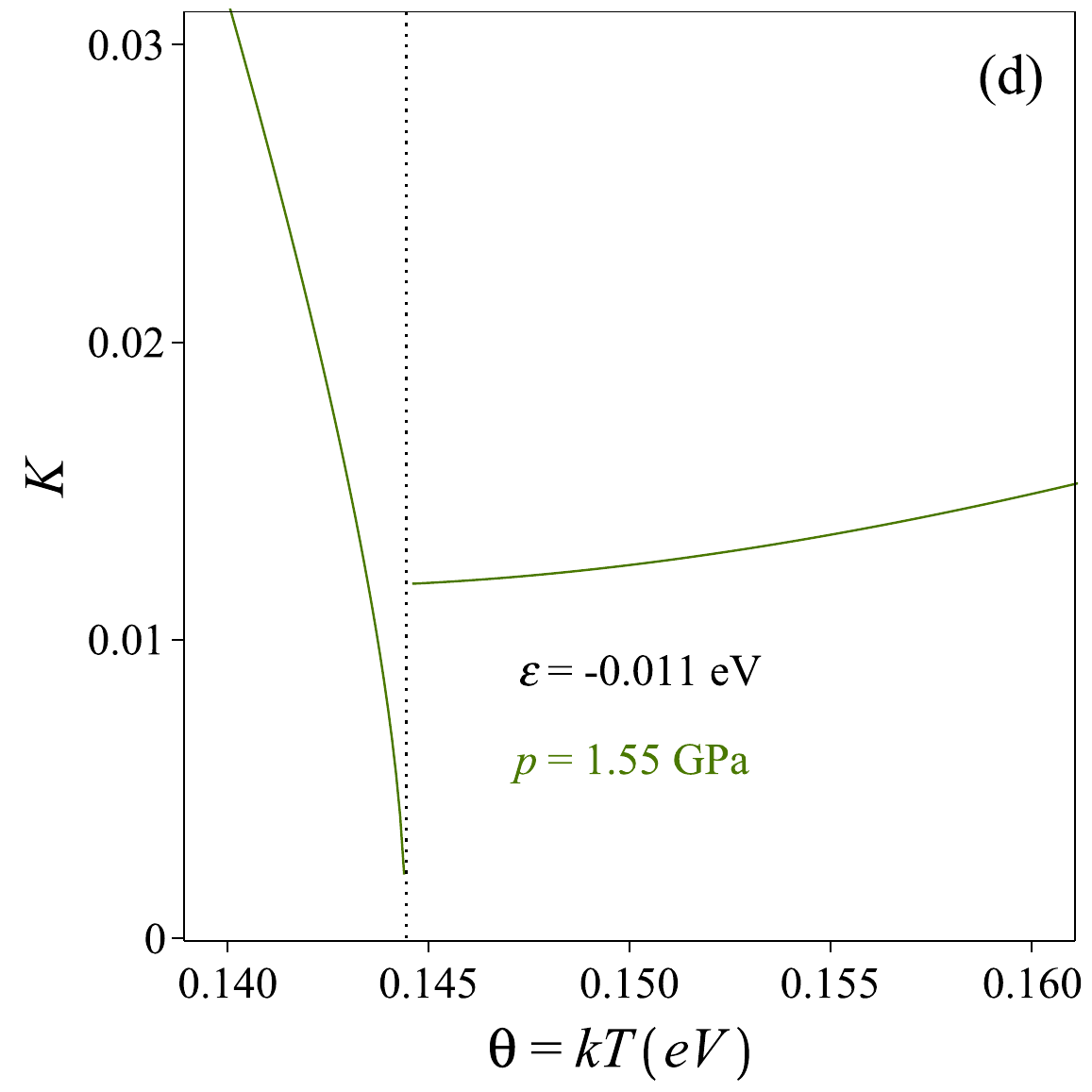}
	\caption{(Colour online) \textbf{(a)}, \textbf{(b)} $H$ and \textbf{(c)}, \textbf{(d)} $K$ vs. $\theta$ for several pressures  at  $\varepsilon=-0.011$~eV.} \label{fig3}
\end{figure*} 

In subsequent two figures (figure~\ref{fig2} and figure~\ref{fig3}), we have plotted $H$ and $K$ as functions of pressure and temperature for the FE and PE solutions to show their critical and tricritical properties. As can be seen from both figures, almost the same results as figure~\ref{fig1} are found. Different from figure~\ref{fig1}(a), the blue curve in figure~\ref{fig2}(a) appears smoother than the red curve in the entire range of $p$, including $p \geqslant p_C$. Furthermore, the finite jump of $H$ $(\Delta H = 0.027)$ at the first-order phase transition $(p_t)$ from the FE phase to PE phase accompanied by the compression of the lattice [figure~\ref{fig2}(b)] is greater than that in figure~\ref{fig1}(b)  $(\Delta H = 0.018)$. As opposed to $K$ in figures~\ref{fig1}(c) and \ref{fig1}(d), the curvature $K$ shown by blue, red and green colored curves in figures~\ref{fig2}(c) and \ref{fig2}(d) displays a maximum at a pressure value (colored arrows) in the PE phase regime. Moreover, the amount of the jump in $K$ $(\Delta K)$ [figure~\ref{fig2}(d)] is less than that in figure~\ref{fig1}(d). Similarly, in figures~\ref{fig3}(b) and \ref{fig3}(d), it is seen that the jump amounts of $H$ and $K$ are slightly less than those  in the previous figures.

\begin{figure*}[h]
	\centering
	\includegraphics[scale=0.25]{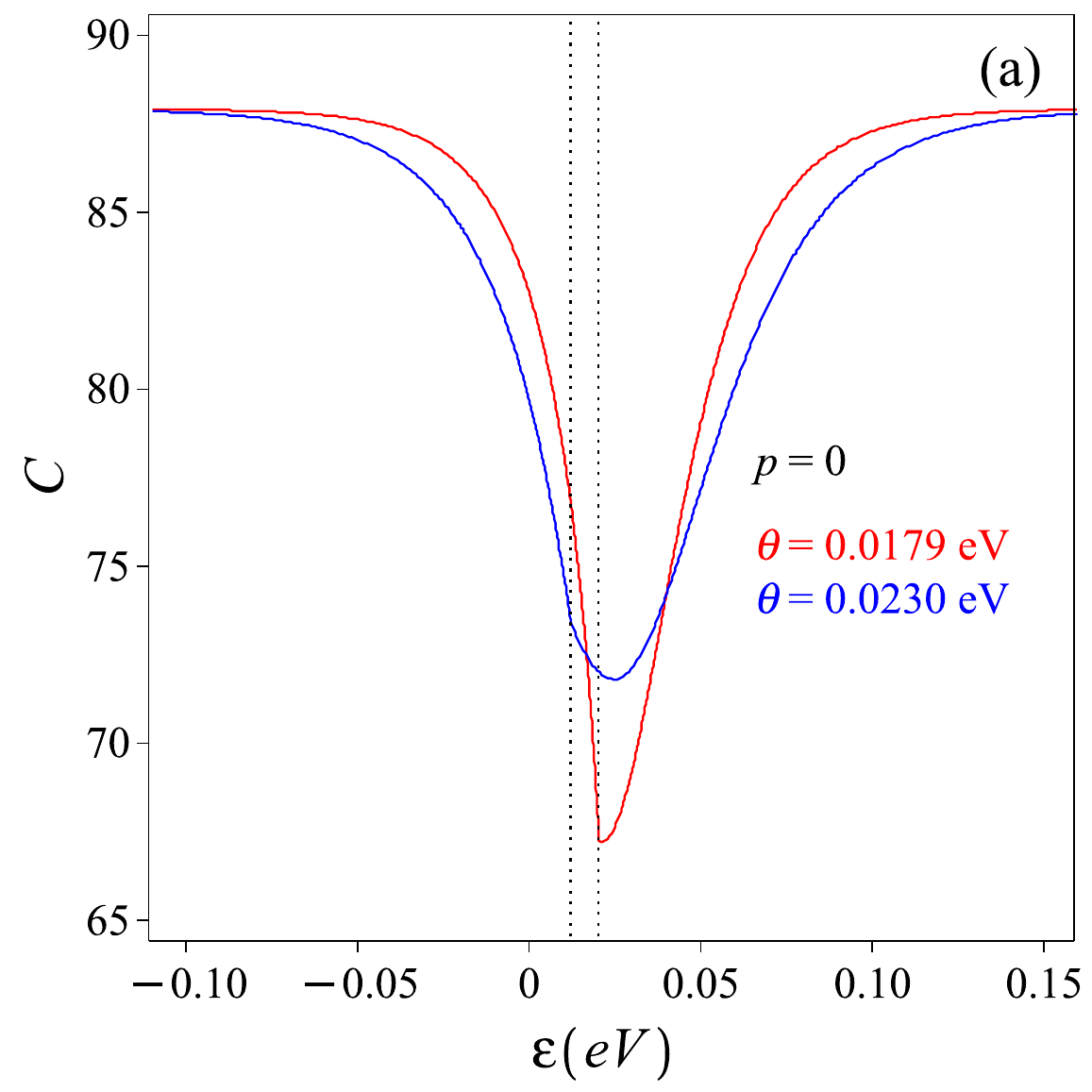}
	\includegraphics[scale=0.25]{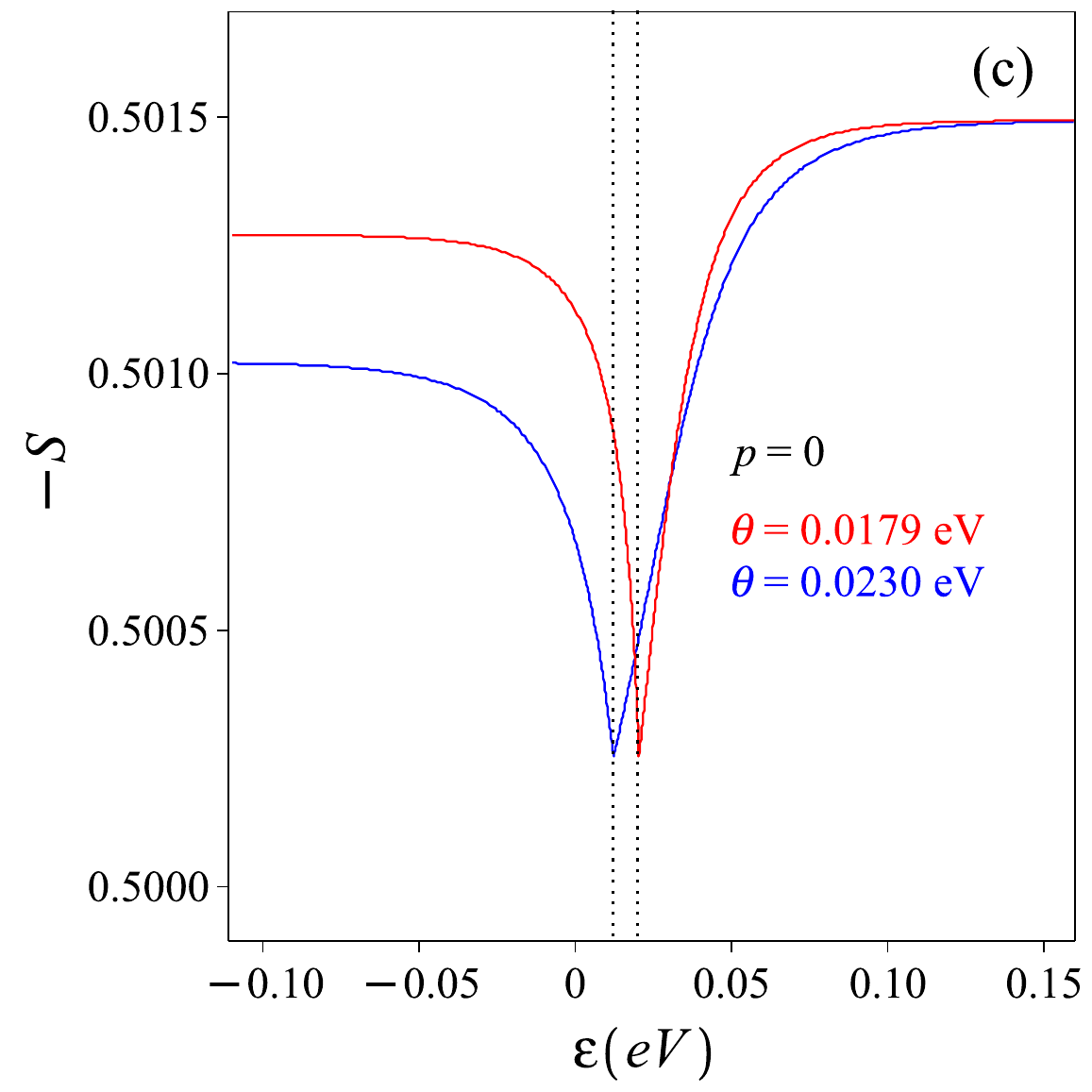}\\
	\includegraphics[scale=0.25]{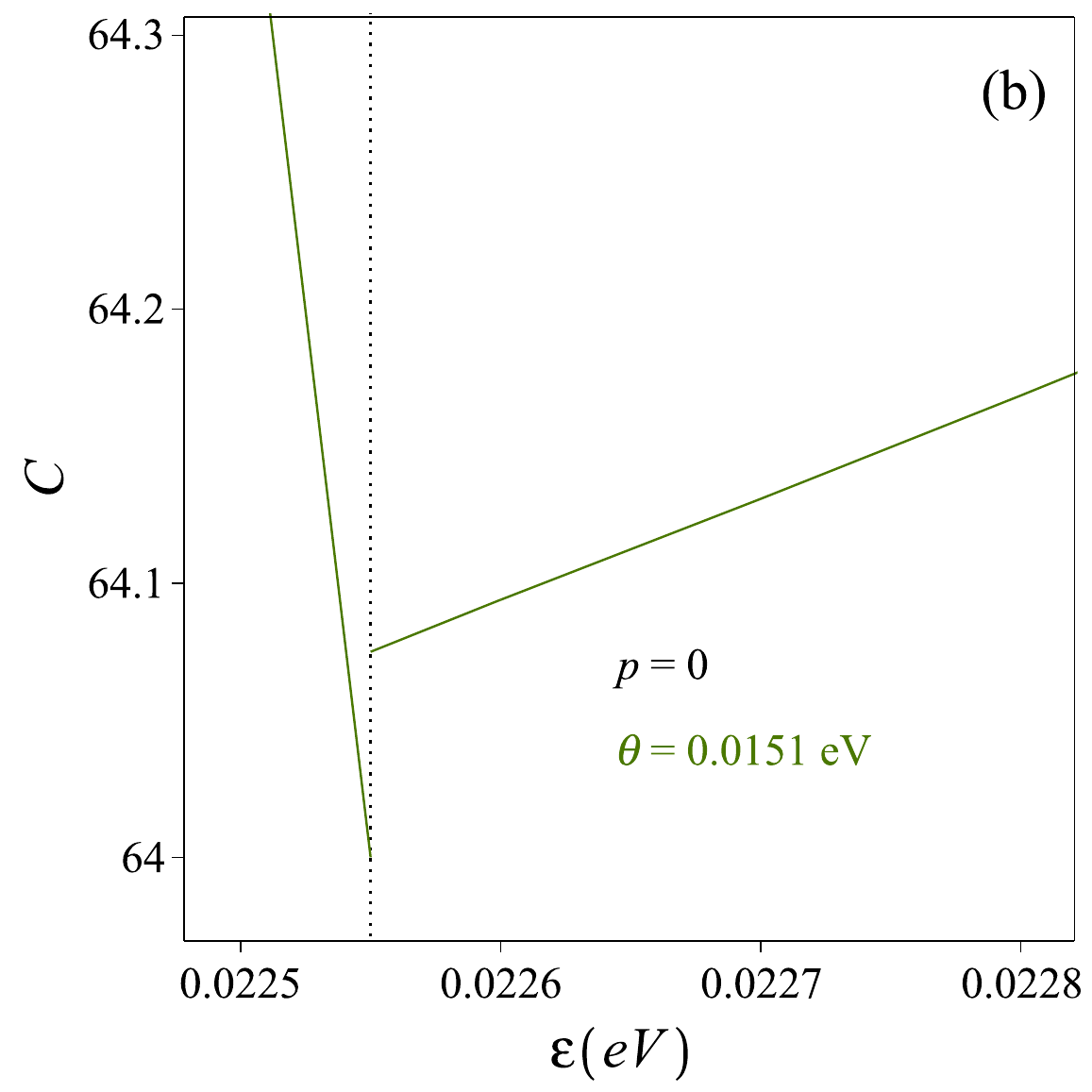}
	\includegraphics[scale=0.25]{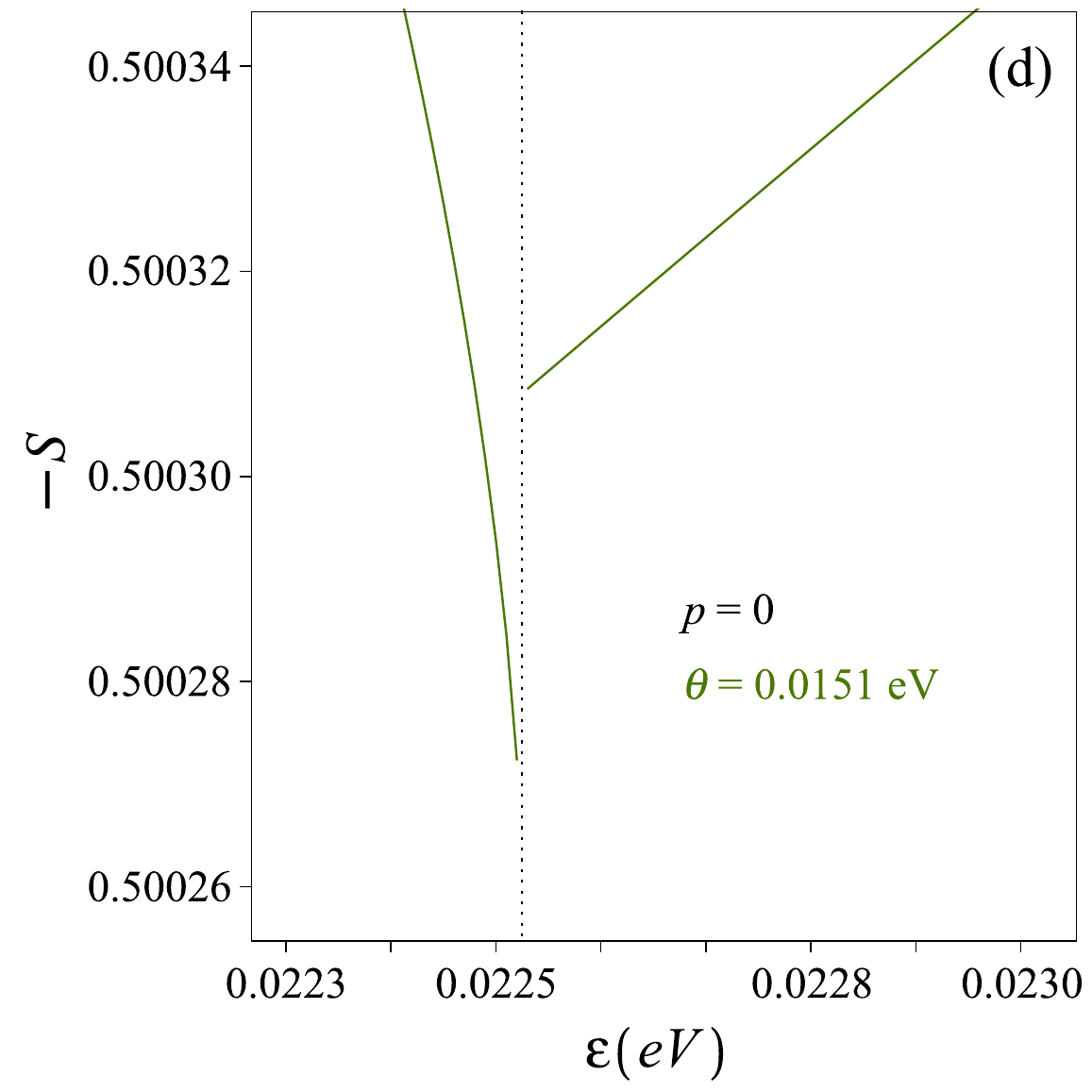}
	\caption{(Colour online) \textbf{(a)}, \textbf{(b)} $C$ and \textbf{(c)}, \textbf{(d)} $S$ vs. $\varepsilon$ for several temperatures  at  $p=0$.} \label{fig4}
\end{figure*}

\begin{figure*}[h!]
	\centering
	\includegraphics[scale=0.25]{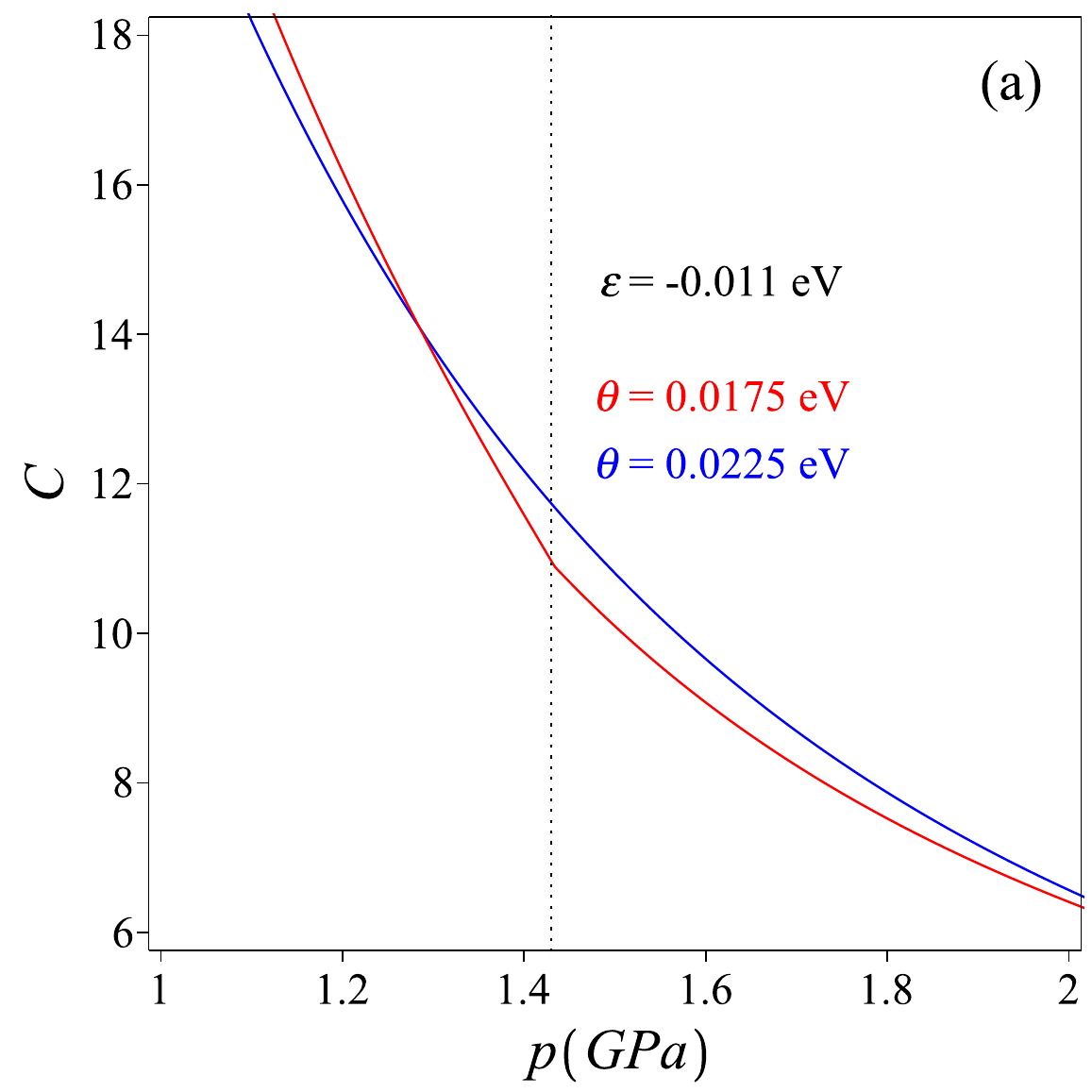}
	\includegraphics[scale=0.25]{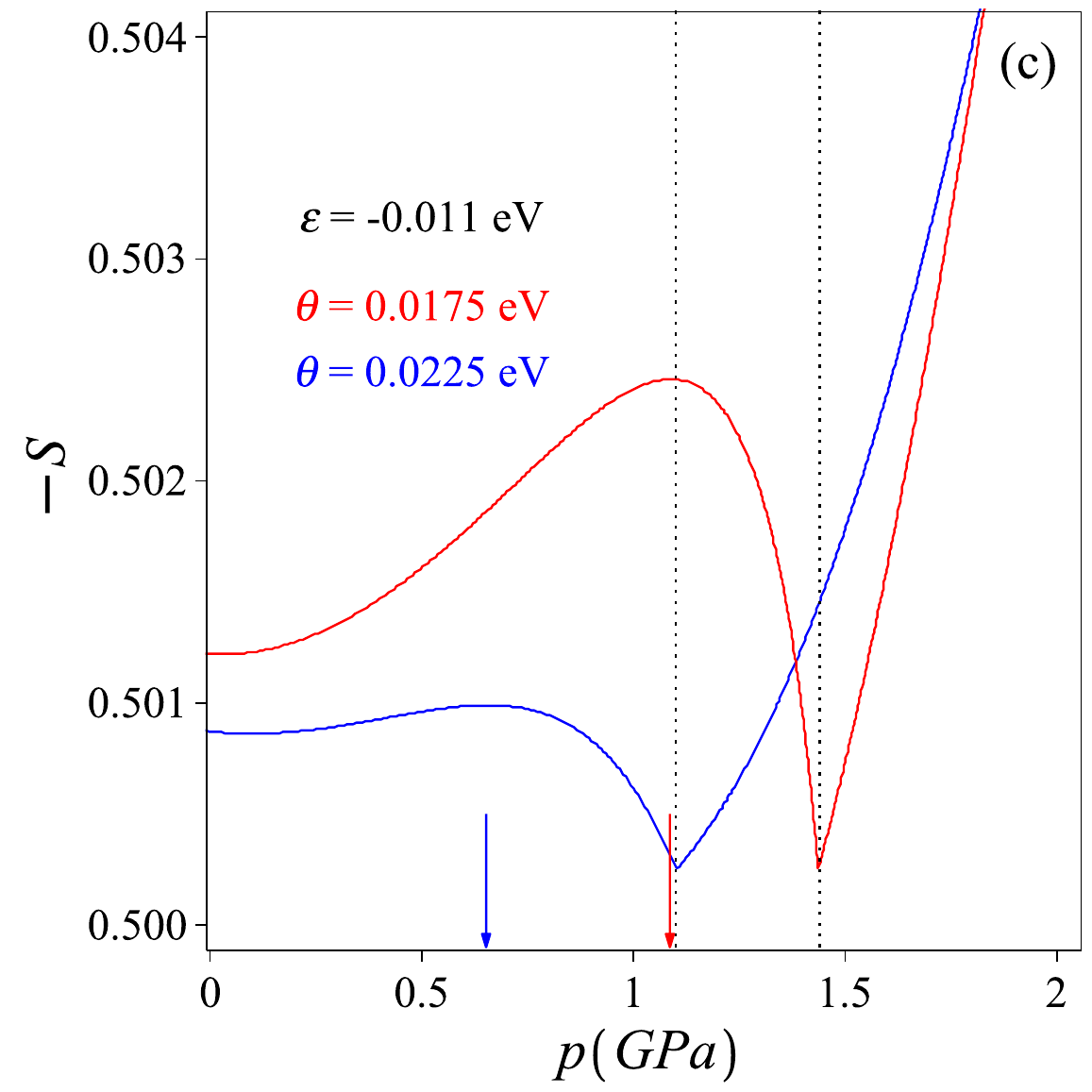}\\
	\includegraphics[scale=0.25]{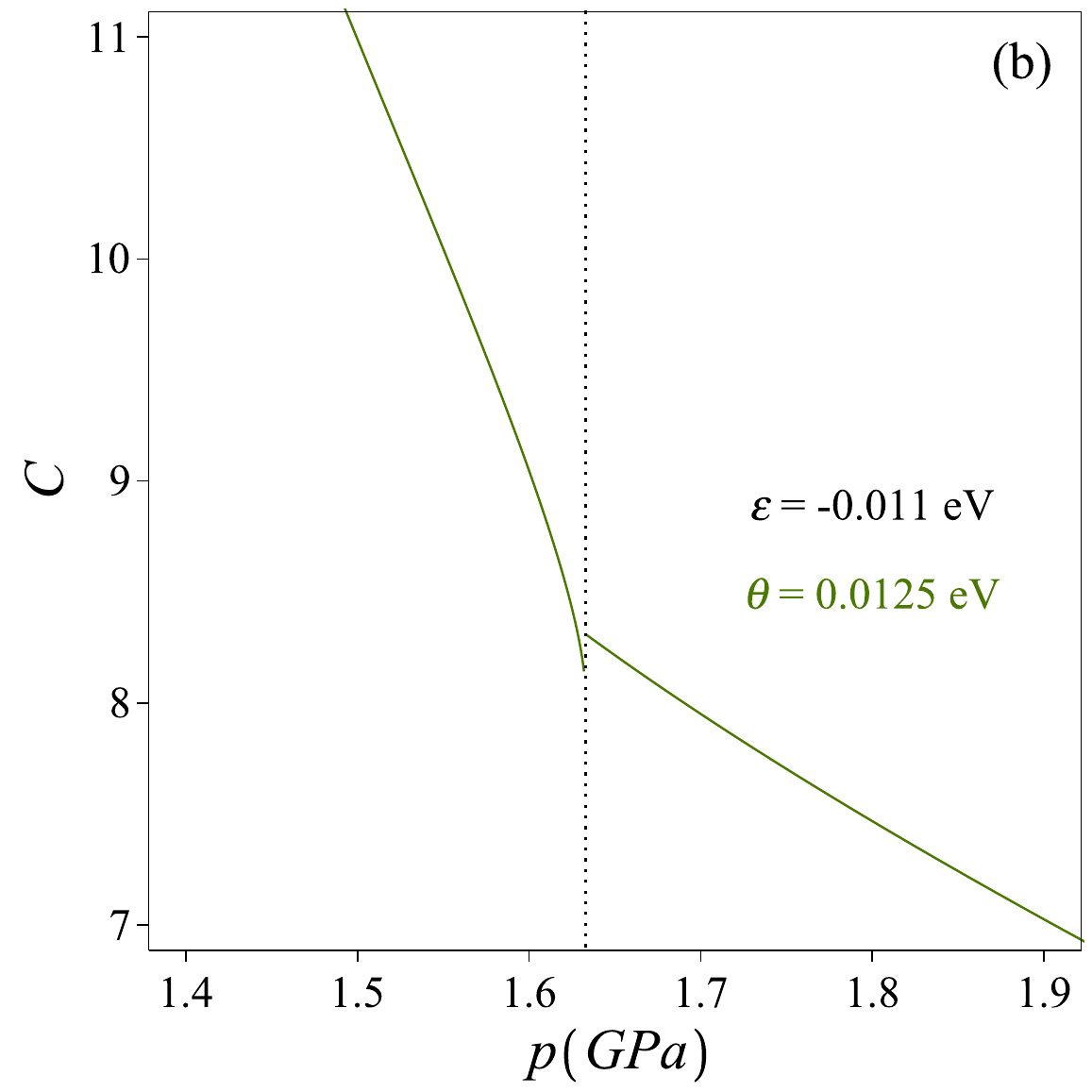}
	\includegraphics[scale=0.25]{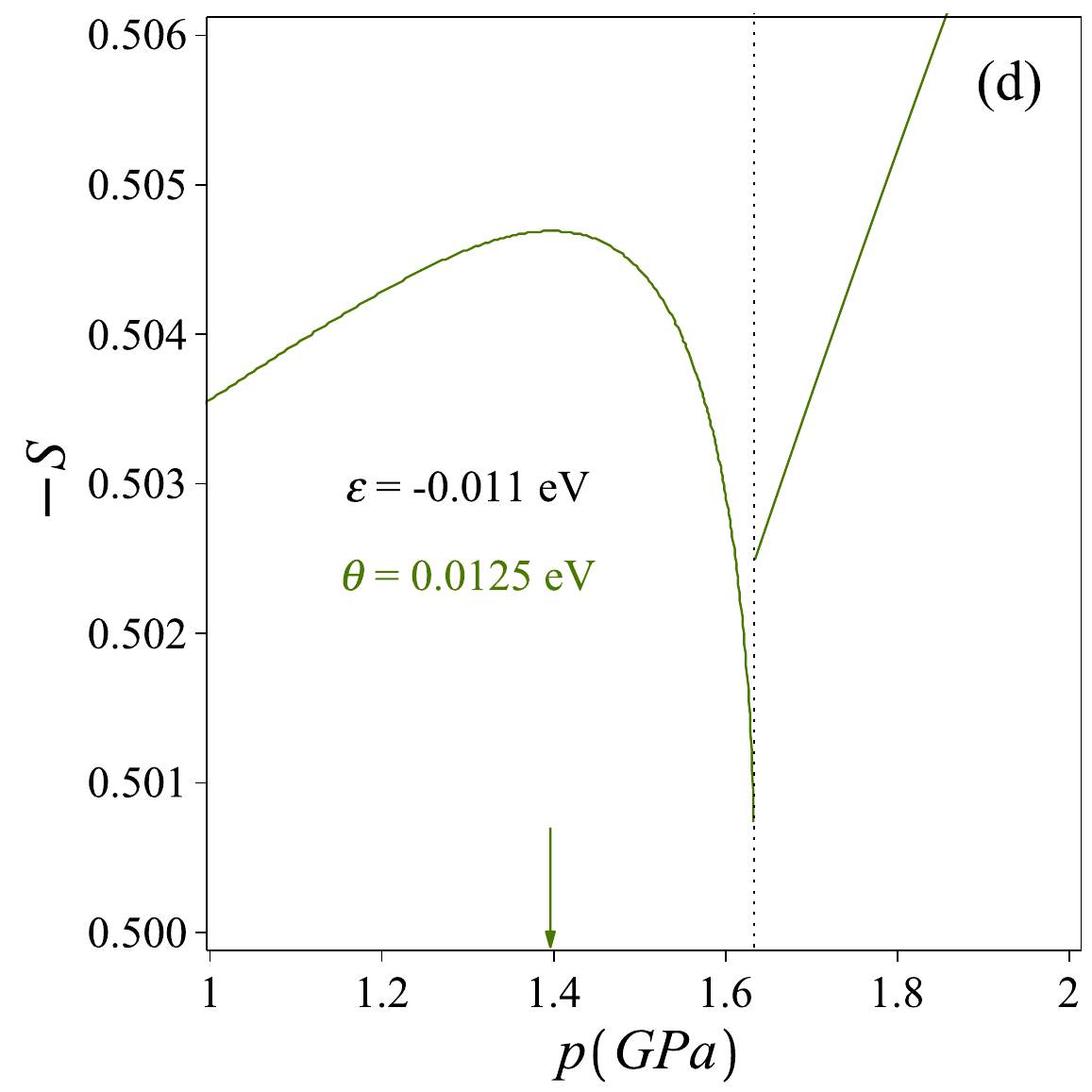}
	\caption{(Colour online) \textbf{(a)}, \textbf{(b)} $C$ and \textbf{(c)}, \textbf{(d)} $S$ vs. $p$ for several temperatures  at $\varepsilon=-0.011$~eV.} \label{fig5}
\end{figure*}

\begin{figure*}[h!]
	\centering
	\includegraphics[scale=0.25]{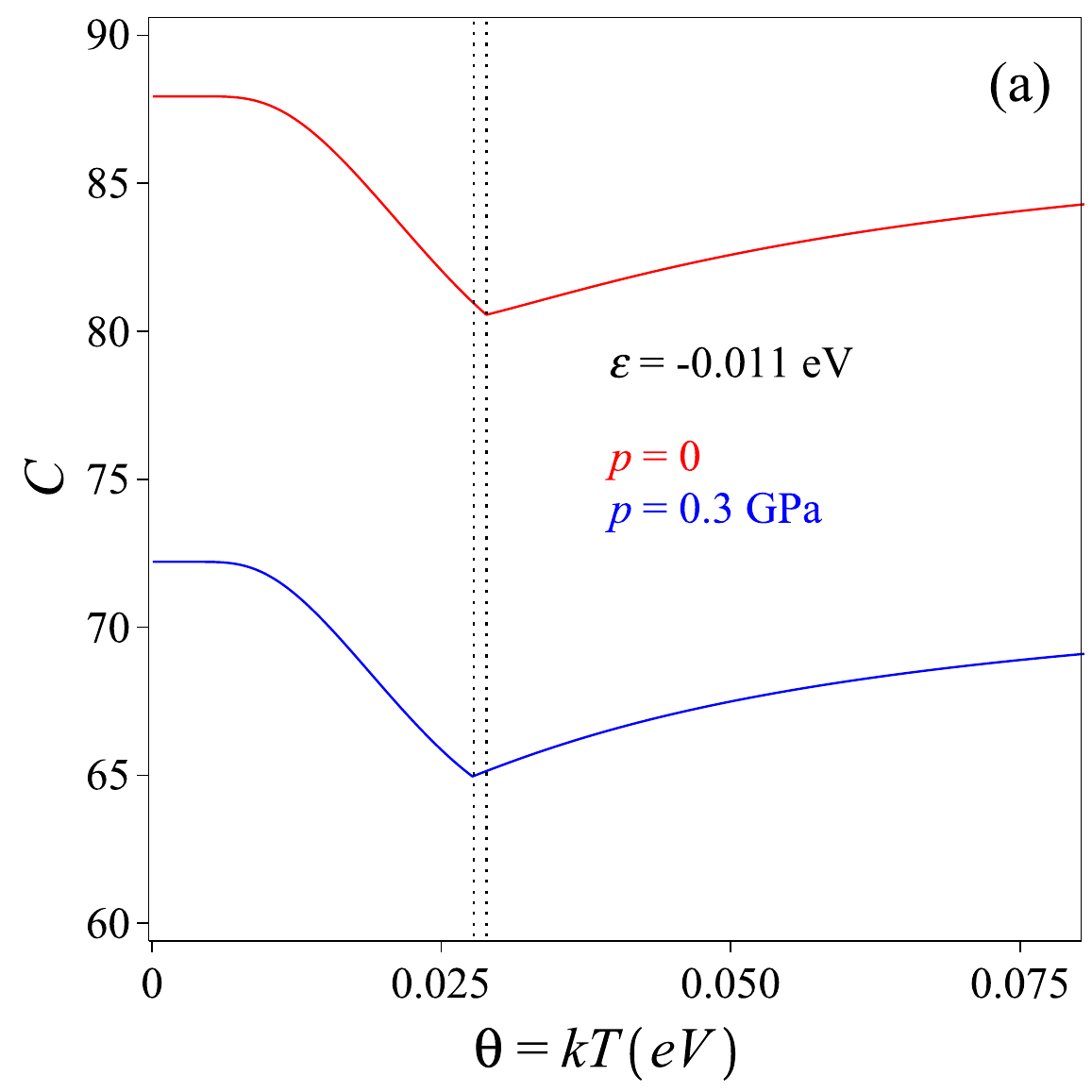}
	\includegraphics[scale=0.25]{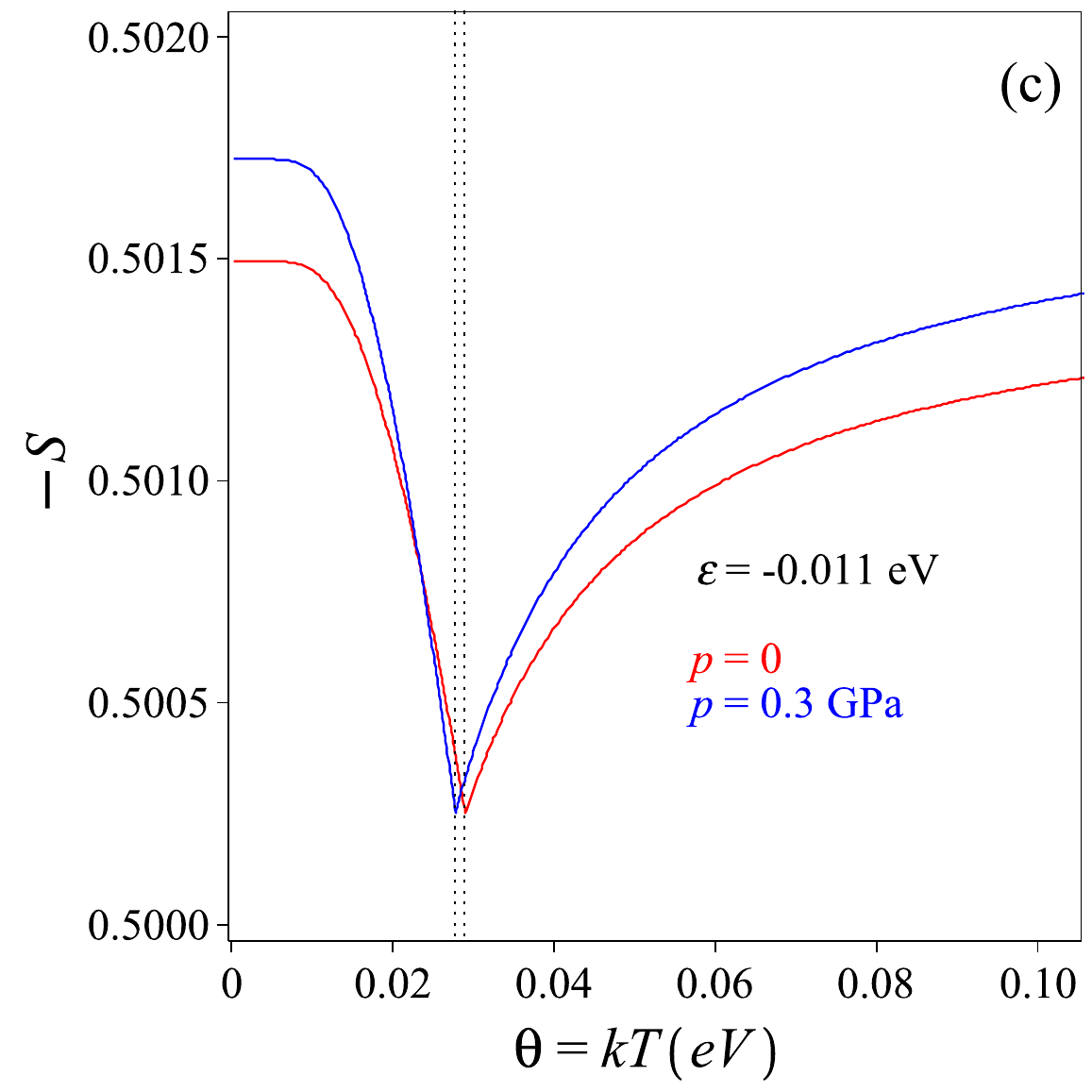}\\
	\includegraphics[scale=0.25]{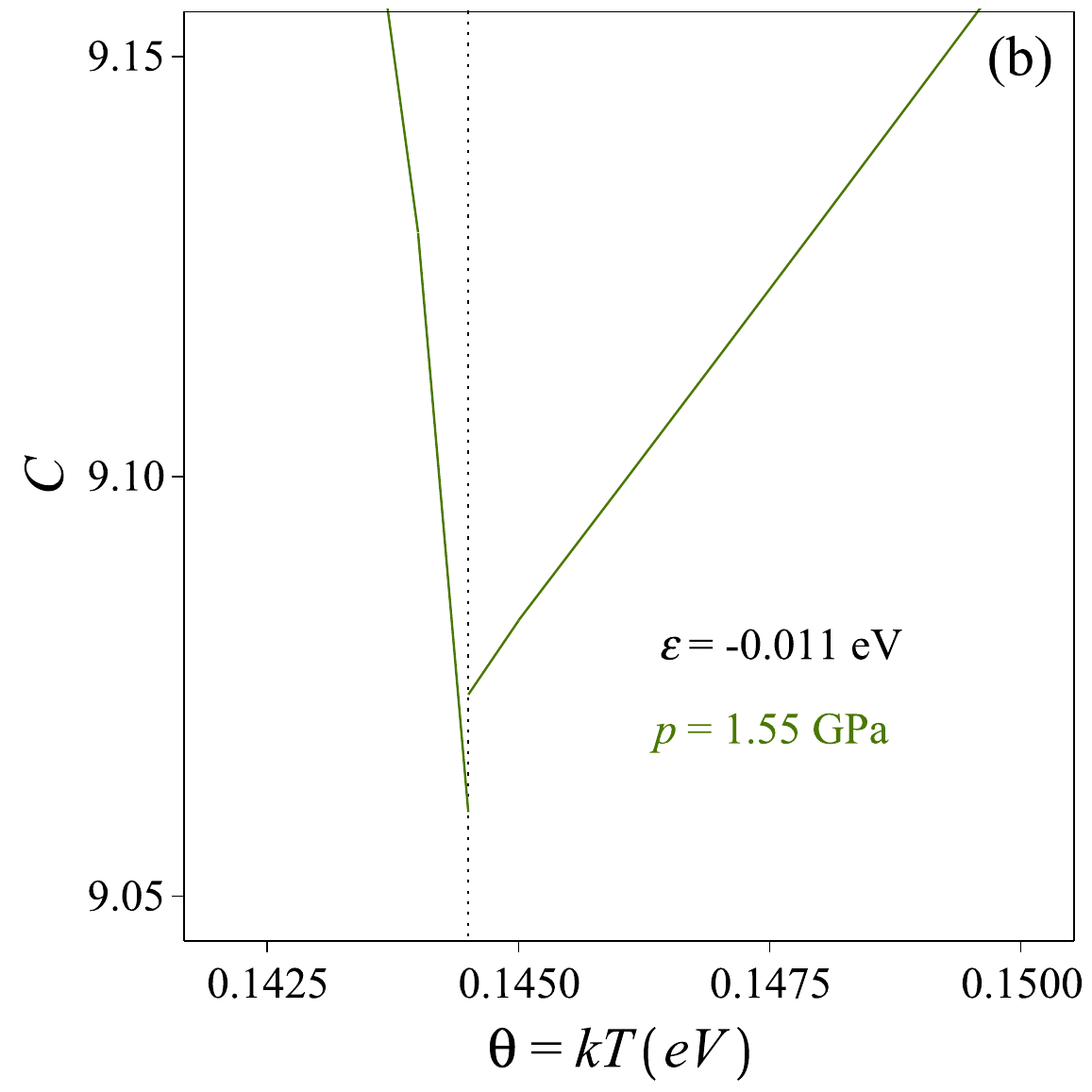}
	\includegraphics[scale=0.25]{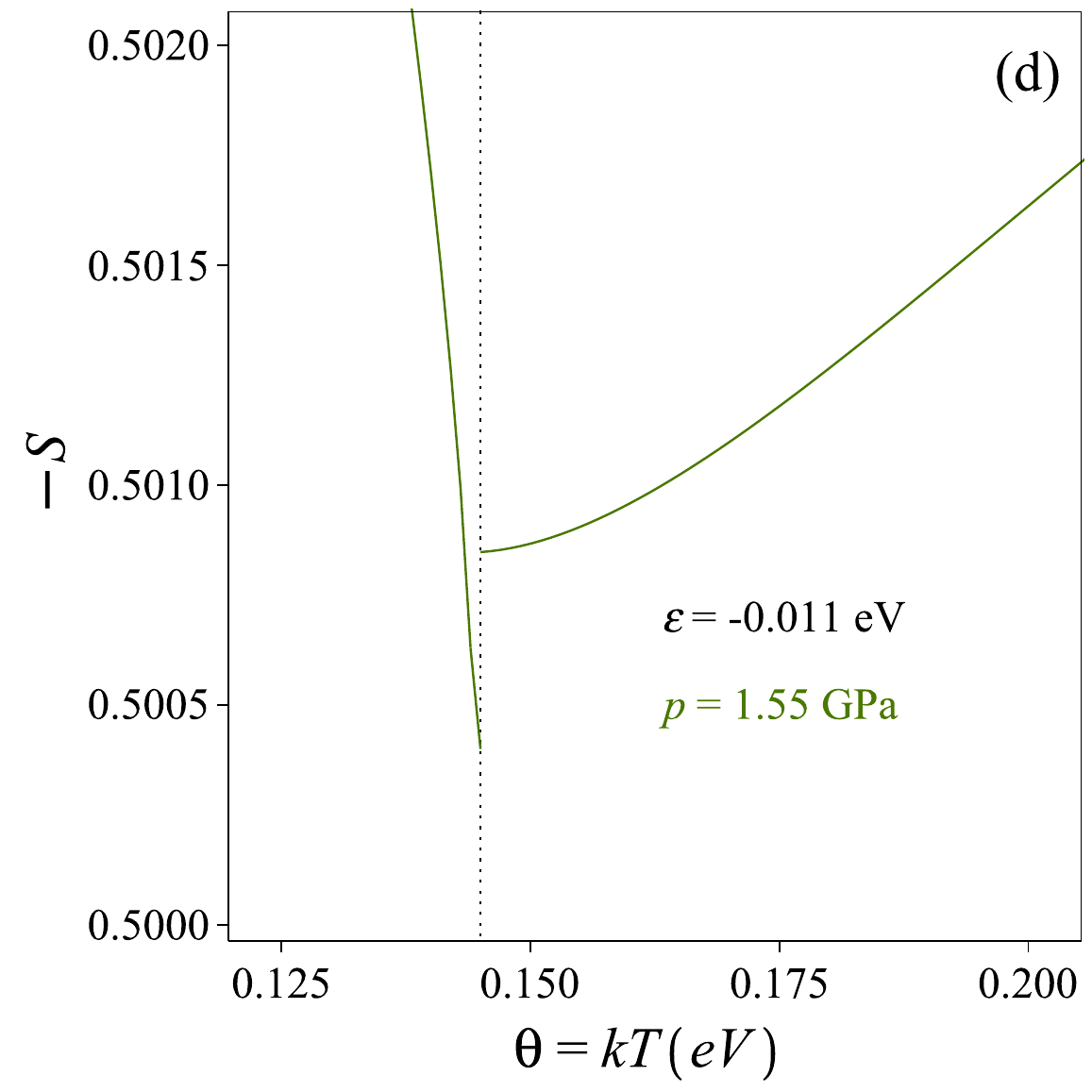}
	\caption{(Colour online) \textbf{(a)}, \textbf{(b)} $C$ and \textbf{(c)}, \textbf{(d)} $S$ vs. $\theta$ for several pressures at $\varepsilon=-0.011$~eV.} \label{fig6}
\end{figure*}

As for the analysis using the curvedness and shape index,  the obtained numerical results are shown in figures~\ref{fig4}, \ref{fig5}, \ref{fig6}. Firstly, using the same temperature values in figure~\ref{fig1} we have illustrated $C$ vs. $\varepsilon$ and $S$ vs. $\varepsilon$ plots in figure~\ref{fig4}. We note here that $C>0$ and $S<0$ in the whole range of $\varepsilon$ values including $\varepsilon \approx \varepsilon_C$ and $\varepsilon \approx \varepsilon_{\rm {TCP}}$. Hence, according to the second category in table~\ref{table-1}, we confirm from figures~\ref{fig4}(a) and \ref{fig4}(c) that the free energy surfaces  are of ``valley shape'' around $\varepsilon_C$ and $\varepsilon_{\rm {TCP}}$. We can also see the same characteristics of $C$ and $S$ at the first-order phase transition point [figures~\ref{fig4}(b) and \ref{fig4}(d)] as those of $H$ and $K$ given in figures~\ref{fig1}(b) and \ref{fig1}(d), respectively. In other words, finite jumps of $C$ and $S$ occur at $\varepsilon_t$.

Secondly, we have studied the pressure variation of  $C$ and $S$  for $\varepsilon=-0.011$ eV in figure~\ref{fig5}, where cusps  at the pressure values $p_C$, $p_{\rm {TCP}}$ and jumps at $p_t$ are also observed as expected. Among the results in the figure, an extremum of $S$ is found in the FE phase regime, indicated by the colored arrows in figures~\ref{fig5}(c) and \ref{fig5}(d). Finally, the dependence of $C$ and $S$  on the temperature $\theta$   can be followed in figure~\ref{fig6}. As a critical behaviour it only displays a cusp singularity at the temperatures $\theta_C, \theta_{\rm {TCP}}$  [figures~\ref{fig6}(a) and~\ref{fig6}(c)], while one observes a jump discontinuity of  $C$ and $S$   at $\theta_t$ [figures~\ref{fig6}(b) and~\ref{fig6}(d)]. It is also quite evident from figure~\ref{fig6}(a) that the investigated temperature plots of $C$ are in direct agreement with $H$ vs. $\theta$ plots in figure~\ref{fig3}(a). On the other hand, by a comparison of the colored curves in figure~\ref{fig6}(c) it can be stated that around the critical temperature $S$  is not much sensitive to any change in  $p$ values.

\begin{figure*}[h!]
	\centering
	\includegraphics[scale=0.25]{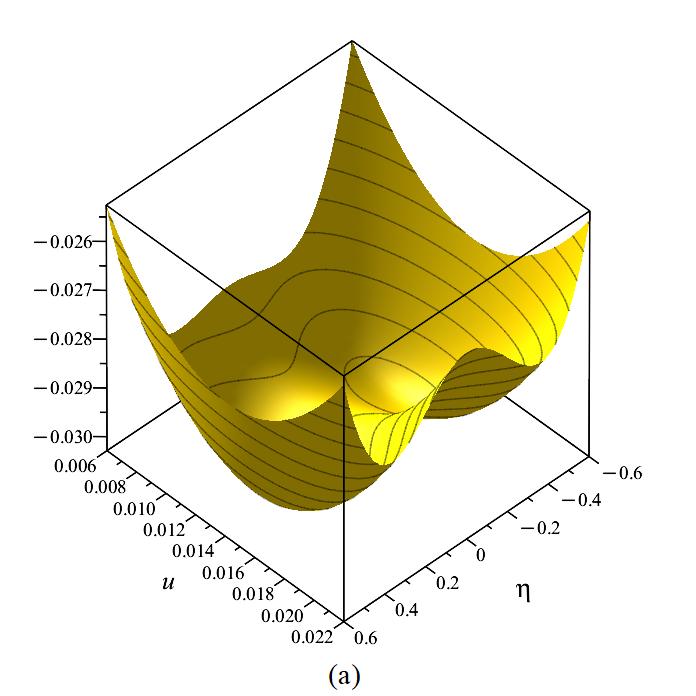}
	\includegraphics[scale=0.25]{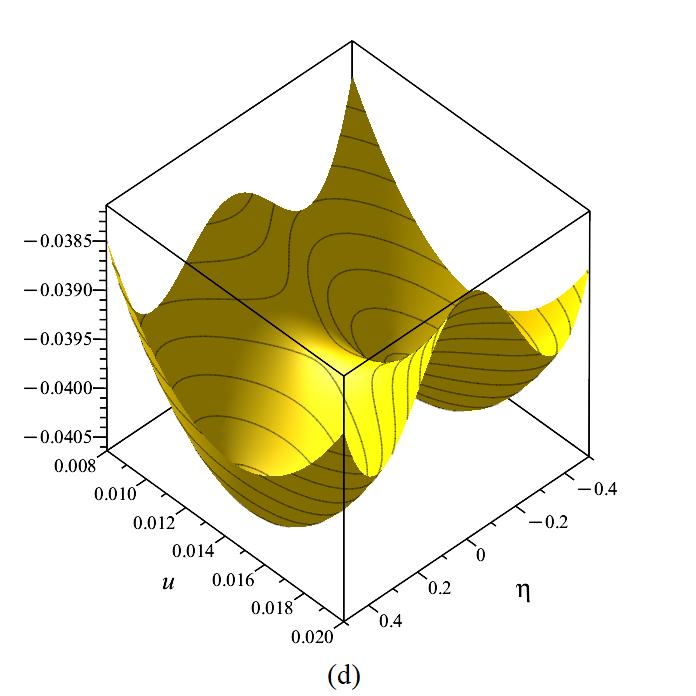}
	\includegraphics[scale=0.25]{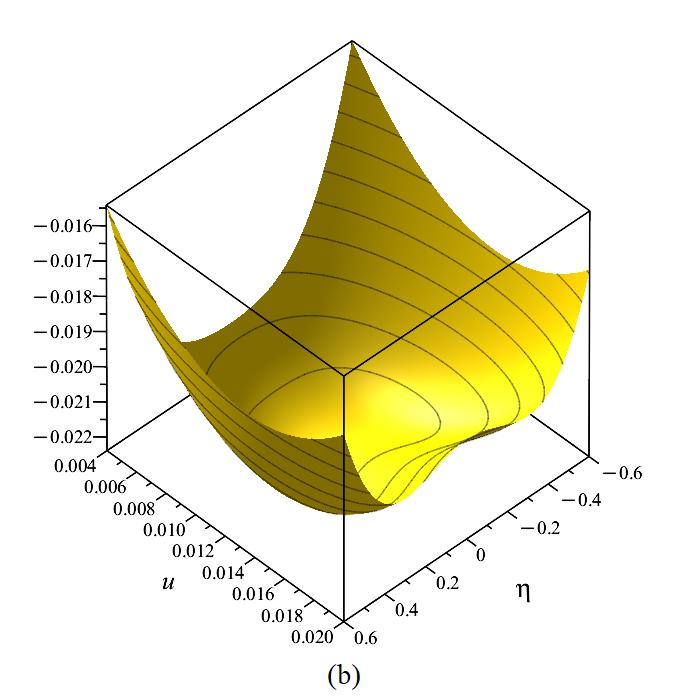}
	\includegraphics[scale=0.25]{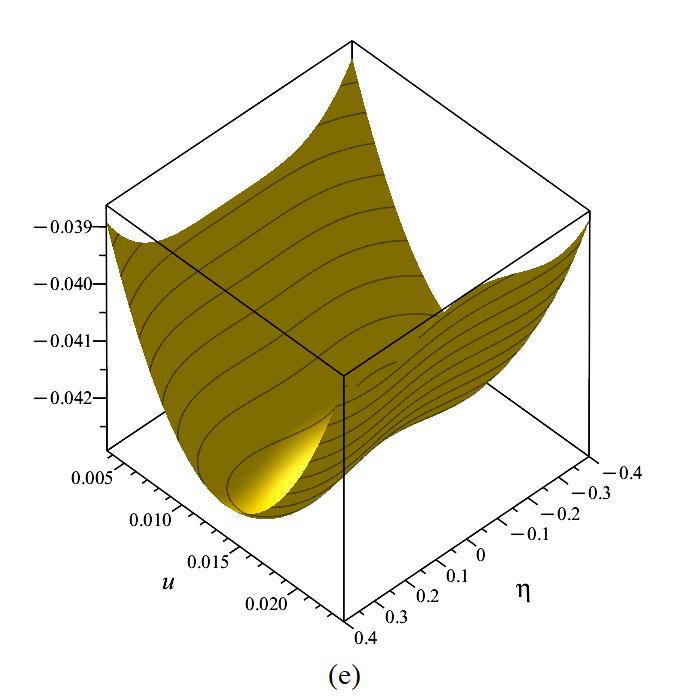}	
	\includegraphics[scale=0.25]{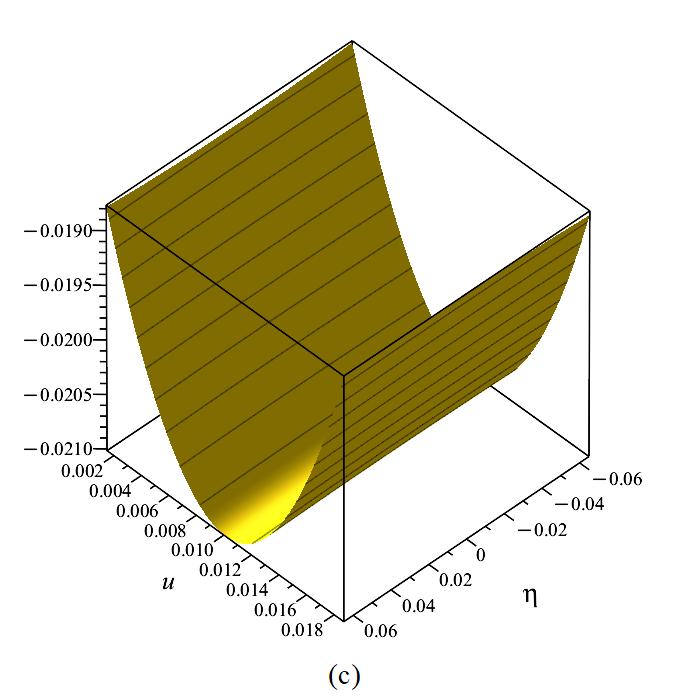}
	\includegraphics[scale=0.25]{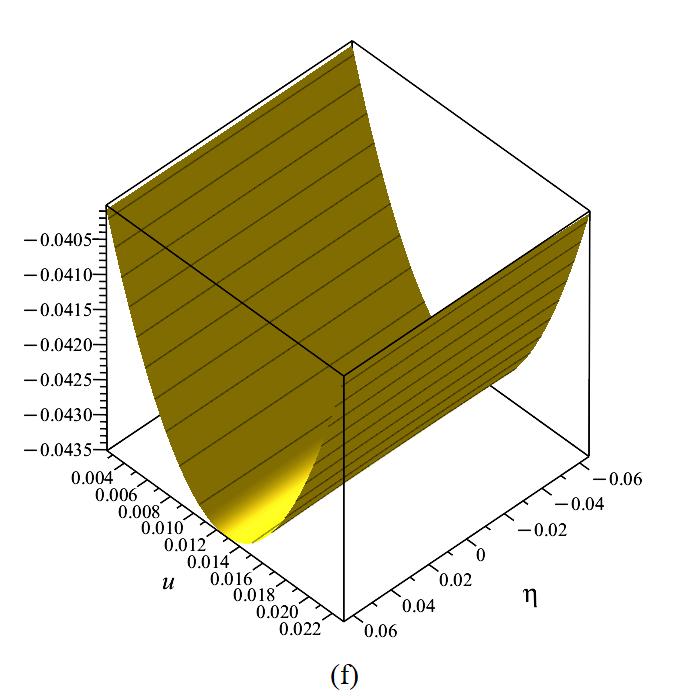}
	\caption{(Colour online) The landscapes of the free energy surfaces in the FE phase region and around the FE--PE phase transition. Results for $p=0$, $\theta = 0.0230$ eV with \textbf{(a)} $\varepsilon = 0.0$ eV ($<\varepsilon_C$),  \textbf{(b)}~$\varepsilon= 0.01$~eV $(\approx \varepsilon_C)$, \textbf{(c)} $\varepsilon = \varepsilon_C = 0.012$ eV and for  $\varepsilon = -0.011$, $p=0.3$ GPa with \textbf{(d)} $\theta = 0.024$  eV $(<\theta_C)$, \textbf{(e)}~$\theta =  0.027$~eV $(\approx \theta_C)$, \textbf{(f)}~$\theta = \theta_C = 0.0277$  eV.} \label{fig7}
\end{figure*}

In order to confirm and visualize the above properties found in figures~\ref{fig1}--\ref{fig6}, we can easily present the surface plots in $3D$ for a narrow ranges of $\eta$ and $u$. To this end, using $\theta = 0.0230$ eV and some of the $\varepsilon$ values in the absence of pressure $(p = 0)$ and $\varepsilon = -0.011$ eV at various temperatures when $p \neq 0$, we have illustrated the morphology of the thermodynamic potential predicted by our calculations. Among them, four surfaces in the FE phase and two surfaces around the FE--PE phase transition point are shown in figure~\ref{fig7}. As illustrated in figure~\ref{fig7}(a), at $\varepsilon<\varepsilon_C$ case in the $(\eta, u)$ surface which is located between the intervals $-0.6<\eta<0.6$ and $0.006<u<0.022$ on the $\eta$ and $u$ axes two minima (or pits) are seen explicitly predicted in figures~\ref{fig1}(a), \ref{fig1}(c), \ref{fig4}(a) and \ref{fig4}(c). Changes of $\varepsilon$ towards $\varepsilon_C$ ($\varepsilon \approx \varepsilon_C$) causes the localization of two minima approach each other [figure~\ref{fig7}(b)]. At $\varepsilon=\varepsilon_C$, the surface is exactly a valley shape as expected from the above calculations [figure~\ref{fig7}(c)]. Similarly, in figure~\ref{fig7}(d), one also observes two pits at $\theta<\theta_C$ in the intervals $-0.5<\eta<0.5$ and $0.008<u<0.020$ when $p=0.3$ GPa. Upon increasing the temperature, the pits disappear and become a valley, illustrated in figures~\ref{fig7}(e), and~\ref{fig7}(f).

\section{Conclusions}
\label{sec-Con}

In this work, equilibrium states of a QLM were analyzed within shape descriptors ($H$, $K$, $C$, $S$). Using the estimated model parameters for the SPS ferroelectric crystals in \cite{18}, we obtained their energy gap, pressure and temperature variations to determine the geometric contructions of the free energy surfaces in the FE and PE phases. Specifically, we have also shown the behaviours of all descriptors near the continuous and discontinuous phase transitions between these phases. From the figures, it is clear that our calculations present a ``spherical cup'' or ``pit shape'' which corresponds to a stable state in the free energy profile when the system is far from the critical and tricritical points. By contrast, a ``valley'' or ``rut shape'' occurs around the criticality or tricriticaly as expected \cite{11}. Due to the convergence of $K$  towards zero near the critical and tricritical regime, it becomes possible to construct the phase diagrams using the Gaussian curvature. When simultaneously substituting the known values of $(\eta, u)$ at the criticality (or tricriticality) into equation \eqref{eq5} and equating it to zero, a simple expression for the phase boundaries is found for the choices of parameter values. This approach seems easier than solving the self-consistency equations derived by the free energy minimization. Such a relevence between Gaussian curvature and critical curves in the phase diagram has been shown recently in a prominent work on Ising systems \cite{11}.

It is worthwhile to mention that the presented approach is usable not only for the ferroelectrics with two parameters $(\eta, u)$ but also in case the thermodynamic potential contains two order parameters and a deformation. From the mathematical definitions given for any $3D$ morphological changes in \cite{1, 8}, one can easily derive and calculate the above descriptors using a mean-field thermodynamic potential and adapt the presented scheme for a more complex system, such as those with two ferroactive sublattices like CuInP$_2$S$_6$ ferrielectrics \cite{23}. Hence, using the descriptors ($H$, $K$, $C$, $S$), the complete shapes of the free energy surfaces at equilibrium obtained self-consistently can be explained. Particularly, as the critical or tricritical point is approached, one can observe how the unstable states envolve towards metastable or stable states which were previously predicted by a variety of spin-$1$ Ising systems using only the contour mapping methods \cite{24, 25, 26}. Our geometrical interpretation applied previously for the spin systems \cite{11} is now aimed at numerous ferroelectric/ferrielectric systems including metastable and unstable states.

\ukrainianpart

\title{Дескриптори форми рівноважних станів у квантовій ґратковій моделі з локальними багатоямними потенціалами: геометричний аналіз поблизу фазових переходів у сегнетоелектричних кристалах Sn$_2$P$_2$S$_6$}

\author{C. Озум\refaddr{label1},
	Т. Аккурт\refaddr{label2},
	Р. Ердем\refaddr{label3},
	Г. Гючлю\refaddr{label4}}

\addresses{
	\addr{label1} Професійно-технічна школа Алака Авні Челік, Хітітський університет, 19600 Чорум, Туреччина
	\addr{label2} Інститут науки, Університет Акденіз, 07058 Анталія, Туреччина
	\addr{label3} Фізичний факультет, Університет Акденіз, 07058 Анталія, Туреччина
	\addr{label4} Департамент фізичної освіти, Університет Неджметтіна Ербакана, 42090 Конья, Туреччина}

\makeukrtitle

\begin{abstract}
	\tolerance=3000%
	Ми аналізуємо рівноважні стани квантової ґратки з локальними багатоямними потенціалами для сегнето\-електричних кристалів Sn$_2$P$_2$S$_6$, використовуючи середню та гаусову кривини ($H$, $K$), кривину ($C$) та індекс форми ($S$). На основі змін енергетичної щілини, тиску та температури $H$, $K$, $C$ та $S$ ми провели геометричну побудову поверхонь вільної енергії для сегнетоелектричної та параелектричної фаз. Їхня поведінка спостерігається явним чином поблизу фазових переходів ``сегнетоелектрик-параелектрик''. Помічено, що $H$, $C$ та $S$ мають сильну сингулярність у критичній точці, тоді як $K$ прямує до нуля по обидва боки від критичної та трикритичної точок.
	\keywords {середня та гауссова кривизни}, {вигнутість}, {індекс форми},  {квантова ґраткова модель},  {сегнетоелектричні кристали},   {фазові переходи}
	
\end{abstract}

\lastpage
\end{document}